\documentclass{aa}  

\usepackage{graphicx}
\usepackage{txfonts}

\begin{document}

   \title{Waiting times between gamma-ray flares of Flat Spectrum Radio Quasars, and constraints on emission processes}
   \titlerunning{Waiting times between gamma-ray flares of FSRQ}

   \author{L. Pacciani \inst{1} }

   \institute{Istituto di Astrofisica e Planetologia Spaziali - Istituto Nazionale di Astrofisica (IAPS-INAF),
              Via Fosso del Cavaliere, 100 - I-00133 Rome (Italy)\\
              \email{luigi.pacciani@inaf.it} }

\authorrunning{L. Pacciani \inst{1}}

 \abstract
     {The physical scenario responsible for gamma-ray flaring activity and its location for Flat Spectrum Radio Quasars (FSRQ) is still debated.}        
       {The study of the statistical distribution of waiting times between flares (defined as the time intervals between consecutive activity peaks,
       \citealp[see][for a formal description]{wheatland2002})
       can give information on the distribution of flaring times, and constrain the
       physical mechanism responsible for gamma-ray emission.}
     {
     We adopt here a Scan-Statistic driven clustering method
     (\emph{iSRS}, \citealt{pacciani2018})
     to recognize flaring states within the FERMI-LAT archival data, and identify the time of activity peaks.
   }
   {
     We obtained that flares waiting times can
     be described with a poissonian process, consisting of a set of overlapping bursts of flares, with an average burst duration
     of $\sim$0.6 year, and average rate of $\sim$1.3y$^{-1}$.\\    
     For short waiting times (below 1d host-frame) we found a statistically relevant second population, the fast-component,
     consisting of a few tens of cases, most of them revealed for CTA 102.
     Interestingly, the period of conspicuous detection of the fast component of waiting-times for CTA 102 coincides
     with the crossing time of the superluminal K1 feature with the C1
     stationary feature in radio reported in \citet{jorstad2017,casadio2019}.
%
   }
   {
     To reconcile the mechanism proposed in \citet{jorstad2017,casadio2019} with the bursting activity,
     we have to assume that plasma streams with a typical
     length of  $\sim$2pc (in the stream reference frame) reach the recollimation shock.\\
     Otherwise, the distribution of waiting times can be interpreted as 
     originating from relativistic plasma moving along the jet for a deprojected length of $\sim$30-50pc
     (assuming a bulk $\Gamma$=10), that sporadically produces gamma-ray flares.\\
%
%
     In  magnetic reconnection scenario,
     reconnection events or plasma injection to the reconnection sites should be intermittent.
     Individual plasmoids can be resolved in a few favourable cases only \citep{christie2019};
     they could be responsible for the fast component.\\
   }




\maketitle

\section{Introduction}
Blazars are the most luminous extragalactic objects observed in
gamma-ray. They emit from Radio to TeV.\\
There is a general consensus that they are among the active galactic nuclei
(AGN) that are able to produce a jet approximately oriented  along the polar
axis.
In particular, blazars are jetted AGN for which the line of sight of the
observer lies close to the jet axis \citep{urry1995}.\\
VLBA images of AGN jets at 15 GHz show that jets geometry in the region from 10$^2$ to 10$^3$ pc from the core is approximately
conical with an aperture of $\sim1^\circ$ \citep{pushkarev2017}.
Several jets have parabolic shape on shorter distances.\\
Relativistic features \citep{lister2016,hovatta2009} moving downstream the jet are observed.
They do not fill the entire cross-section of the ejection cone \citep{lister2013}.
Their motion is not always ballistic, showing radial and non-radial acceleration.\\ 
%
%
According to the Broad-Line emission power with respect to the optical
continuum  \citep{urry1995}, blazar are sub-divided in Flat Spectrum Radio Quasar (FSRQ, showing
strong Broad Emission Lines) and BL Lac objects (with weak or absent Broad Emission lines).\\
The gamma-ray observed from blazars is due to the Doppler boosted emission from the jet.\\
The mechanism responsible for gamma-ray emission is still a debated issue.
It is also controversial the kind of accelerated particles (leptons, hadrons) originating
the gamma-ray emission, the accelerating engine, and  its location.\\ 
Particles can accelerate through the shock diffusive acceleration \citep{aller1985,hughes1985,blandford1987}.
In this model, piston-driven shock originates blazar outbursts; the shock compresses and orders the upstream
magnetic field, causing the electron to accelerate though the Fermi mechanism, and the  emitted
synchrotron radiation to be polarized.\\  
Alternatively, \citet{narayan2012,marscher2013,marscher2014} proposed turbulence inside the
jet as accelerating mechanism: flares 
are triggered by the passage of turbulent relativistic plasma flow in a
re-collimation shock, located at parsec scale from the central supermassive black hole (SMBH).
The shock  compresses  the plasma and accelerates
electrons. The emission from single turbulent cells is responsible for rapid flares.\\
In magnetically dominated flows,
magnetic reconnection driven by magnetic instabilities along the jet can efficiently accelerate particles
\citep{zenitani2001,guo2014}.
\citet{giannios2013} found that magnetic reconnection events
lead to the formation of plasmoids, whose emission overlaps in an envelope.
Exceptional plasmoids can grow and produce fast flares that can be resolved from
the envelope . \\
In leptonic models for FSRQs emission, the Inverse-Compton on an external target photon field (EC) is
often adopted to explain the gamma-ray emission from relativistic electrons \citep{maraschi1992}.
Seed photons can be the continuum and line emission from the Broad Line Region, or the black body emission from
dusty torus.\\
There is general agreement that if this is the emission mechanism, the flare peak-luminosity is
proportional to the energy density of the external photon field \citep{dermer1997}.\\
It is also found that electrons radiating via Inverse-Compton mechanism on
external photon field dissipate their kinetic energy with a cooling time which
is inversely proportional to the energy density of the target field \citep[see, e. g., ][]{felten1966}.\\
Alternatively, synchrotron radiation produced during flares is the target photon field
for inverse-Compton scattering \citep[Synchro-Self Compton model, SSC,][]{maraschi1992,marscher1992}.
This model is often invoked to model the gamma-ray emission of BL Lac objects.\\
Protons accelerated within the jet could reach the threshold for photo-pion production.
In this model, the proton-synchrotron, the $\pi_0\rightarrow\gamma\gamma$ decay are responsible for jet
gamma-ray emission \citep[hadronic model, see. e.g.][]{bottcher2013}.\\
%
\citet{casadio2015,casadio2019} showed that several gamma-ray flares of FSRQs can be associated in time
with superluminal features crossing stationary features along the jet
observed at 43 GHz.
They  associated the stationary feature C1 that they observed  in the jet of CTA 102 
with a recollimation shock. In particular, \citet{casadio2019} argued that the interaction of the superluminal structure K1 with recollimation
shock C1 generates gamma-ray flaring activity from the end of 2016 to the beginning of 2017.\\
\citet{larionov2020} discussed a similar scenario for 3C 279 observed from radio to gamma-ray for the years 2008-2018.
They indeed noted that the evaluation of crossing-time of radio steady features by superluminal features have large uncertainties,
thence the association of crossing-time with gamma-ray activity had a large chance to be spurious.\\   
%
%
%
In this paper we study flaring activity in gamma-ray of FSRQ contained in the third Fermi-LAT
catalogue \citep[3FGL, ][]{acero2015}.
In particular, we focus on waiting times between activity peaks, defined as the time interval $\Delta_t$
between two consecutive activity peaks \citep[see][for a description of waiting times]{wheatland2002}.
The study is performed over 9.5 y of data, from the beginning
of the FERMI-LAT operation \citep{atwood2009}, up to the failure of a solar panel
actuator. During this period of whole-sky survey, the monitoring with FERMI-LAT
can be considered continuous, with the satellite scanning each region of the
sky approximately every two-four orbits.
Other authors \citep[see, e.g.,][]{abdo2010,nagakawa2013,sobolewska2014,meyer2019,tavecchio2020} studied blazar variability
in gamma-ray, focusing on gamma-ray light curves, rather than on  waiting time between activity peaks. 
 We, instead, disentangle the investigation of gamma-ray variability in a study of a temporal distribution
of waiting times (this paper), and a study of correlated distribution of peak luminosity and flare duration
(discussed in a forthcoming paper).\\ 
It's useful to report here some useful distribution concerning waiting times:
suppose to have a uniformly distributed sample of events within a period of
duration T; the distribution of $\Delta_t$ between consecutive events inside the period is exponential
$\rho=\frac{1}{\tau}e^{-\frac{\Delta_t}{\tau}}$  (where $\tau$ is the mean
waiting time between consecutive events).
 Moreover, the
time interval $\Delta_t=t_{i+k}-t_{i}$  between the events $i$ and $i+k$ of the time-ordered
list of events is Erlang distributed with probability distribution function
$\rho=\frac{1}{\tau^{k}K!}\Delta_t^{k-1}e^{-\frac{\Delta_t}{\tau}}$.
If, for some reason, we do not detect some event, the observed distribution of
waiting times between detected events deviates from an
exponential distribution. For very short waiting times 
($\Delta_t\rightarrow 0$), indeed,  the exponential trend is restored, because
the contribution of the Erlang distribution for $k>1$  tends to 0.\\
%
%
\section{analysis method}
We investigated the FERMI-LAT archival data, prepared data using the standard Fermi
Science Tools (v10r0p5), and used the PASS8 Response Functions.
We applied standard analysis cuts, and selected SOURCE class
events (evclass=128) within 20$^{\circ}$ from the investigated source using \emph{gtselect}.
We applied a zenith angle cut of 90$^{\circ}$ to reject Earth limb gamma-rays.\\
We identified activity periods of each source
using the \emph{iSRS} clustering method \citep{pacciani2018} applied to 9.5 years
of gamma-ray data.\\
%
It is useful to report some definition and the method used in \cite{pacciani2018}:
For each source
an extraction region is chosen for every event, it
corresponds to the 68\% containment radius (the containment radius depends on Energy and event type).
Gamma-ray events within the chosen extraction region around the source catalogue position are selected.
We can evaluate the cumulative exposure of the instrument to the source 
from the start of the FERMI-LAT operation to the time of arrival of each gamma-ray photon:
gamma-ray events are time tagged, and also (cumulative) exposure tagged. 
Clusters of events are obtained applying the \emph{iSRS} procedure in the cumulative exposure domain, not in time domain
(this is a reasonable aspect because the exposure to the source varies with time).
From the dataset of each source, the set of clusters obtained with the \emph{iSRS} procedure are statistically relevant
(we used a threshold for chance probability $P_{thr}\ =\ 1.3\times\ 10^{-3}$, and a tolerance parameter $N_{tol}$=50).
This set is a tree, called \emph{unbinned light curve}.
The clusters representing the activity peaks are the leaves
of the tree.\\
For each leaf, a peak time, and peak flux are identified.
The branches of the tree ending with a leaf are the detected flares of the unbinned light curve:
detected flares are described by a branch that contains at least the leaf.\\
\begin{figure*}
\includegraphics[width=16 cm]{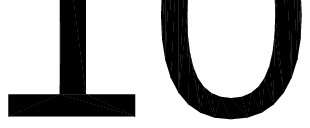} \\
\caption{Unbinned light curve \citep{pacciani2018} for PKS 1510-08, obtained with $P_{thr}=1.3\times 10^{-3}$. Horizontal lines are the statistically relevant clusters.
The set of statistically relevant clusters is a tree, and a hierarchy among clusters exists.
For each son cluster there is a probability $\le\ P_{thr}$ to be obtained starting from its parent cluster by chance.
Vertical bars are  the errors on flux estimate. Background is not subtracted.}
\label{fig_ulc_pks1510}
\end{figure*}
As an example, we report in fig. \ref{fig_ulc_pks1510} the unbinned light curve for PKS 1510-08. The horizontal 
segments are the relevant clusters of events obtained with the \emph{iSRS} procedure. They are described by their starting time, duration,
and by the density of events in the exposure domain, e.g., by the average photometric source flux for the
period of time subtended by the cluster. As far as the background is not subtracted, 
the flux is denoted  with $F_{SRC+BKG}$. Vertical bars represent the error on flux estimate.
Clusters are organized with a tree hierarchy. Each cluster is statistically relevant with respect to
its parent: assuming that events in a parent cluster are uniformly distributed, there is a probability
$\le\ P_{thr}$ that the son cluster is obtained by chance from the parent cluster.
The cluster at the base of the tree is the root.\\
We define \emph{fully resolved} a flare for which the branch contains the leaf and at least
an other cluster, and we define \emph{resolved} a flare for which the branch contains the leaf alone.\\
Photometric flux at peak $F_{phot}^{peak}$ was extracted for resolved flares.
Photometric flux at peak $F_{phot}^{peak}$ and flare duration ($FWHM^{flare}$, full width half maximum)
were extracted for fully resolved flares.\\
We also applied the standard analysis tools to extract flux ($F_{like}^{peak}$, based on likelihood method)
and its significance (from TS statistic)
of the source for the periods of activity identified by the leaves.\\
All photometric methods suffer positional pile-up due to the instrumental Point Spread
Function (PSF) and the presence of nearby sources.
The method adopted here to study source variability is photometric, and it is sensitive to variability of
the positional pile-up (spurious flares). To disentangle spurious flares,  
we classified as spurious, and coming from contiguous sources,
flares with a ratio $R_{photo2like}\ =\ \frac{F_{phot}^{peak}}{F_{like}^{peak}}\ >\ 1.3$, or with $TS\ < $ 9.\\
with this method, we prepared the sample used in this study.
There will remain some contamination in the sample from spurious sources;
and some systematic could affects the results that we will show. We account for the systematic
effects by comparing the results varying the threshold on $R_{photo2like}$ from 1.1 to 2.\\
\section{Samples of activity peaks}
We selected gamma-ray emitting FSRQs, starting from the 3$^{rd}$ FERMI-LAT
catalogue (3FGL).
We identified gamma-ray flaring periods from the FERMI-LAT archival data,
within the period 2008 August - 2018 February.\\
As far as the gamma-ray analysis of sources at low galactic latitude is a cumbersome
task, especially if photometric methods are adopted, we restrict the
investigation to the high latitude FSRQs  ({\em b}$>$ 15$^\circ$) contained
in the third Fermi-LAT catalogue.
To avoid contamination from spurious sources within the 3FGL, we took into
account for bright catalogue sources, with a TS statistic $\ge$ 49.  
With these restrictions, we investigated 335 FSRQs.\\
We obtained two samples of activity peaks, based on the energy range of gamma-ray photons:
A first sample is obtained selecting gamma-ray
events with an energy $E\ >$ 300 MeV (300 MeV sample);\\
Starting from the sources with at least 2 flares in the 300 MeV sample,
we also prepared a sample of activity peaks from the gamma-ray events with $E\ >$ 100 MeV (100 MeV sample).\\
To resolve in time two close flares, large statistics is needed \citep{pacciani2018}. 
In Appendix C, we show the effect on simulated samples.
For this reason,  in this paper we make use of the 100 MeV sample.\\
\section{Results}
\begin{figure*}
\includegraphics[width=16 cm]{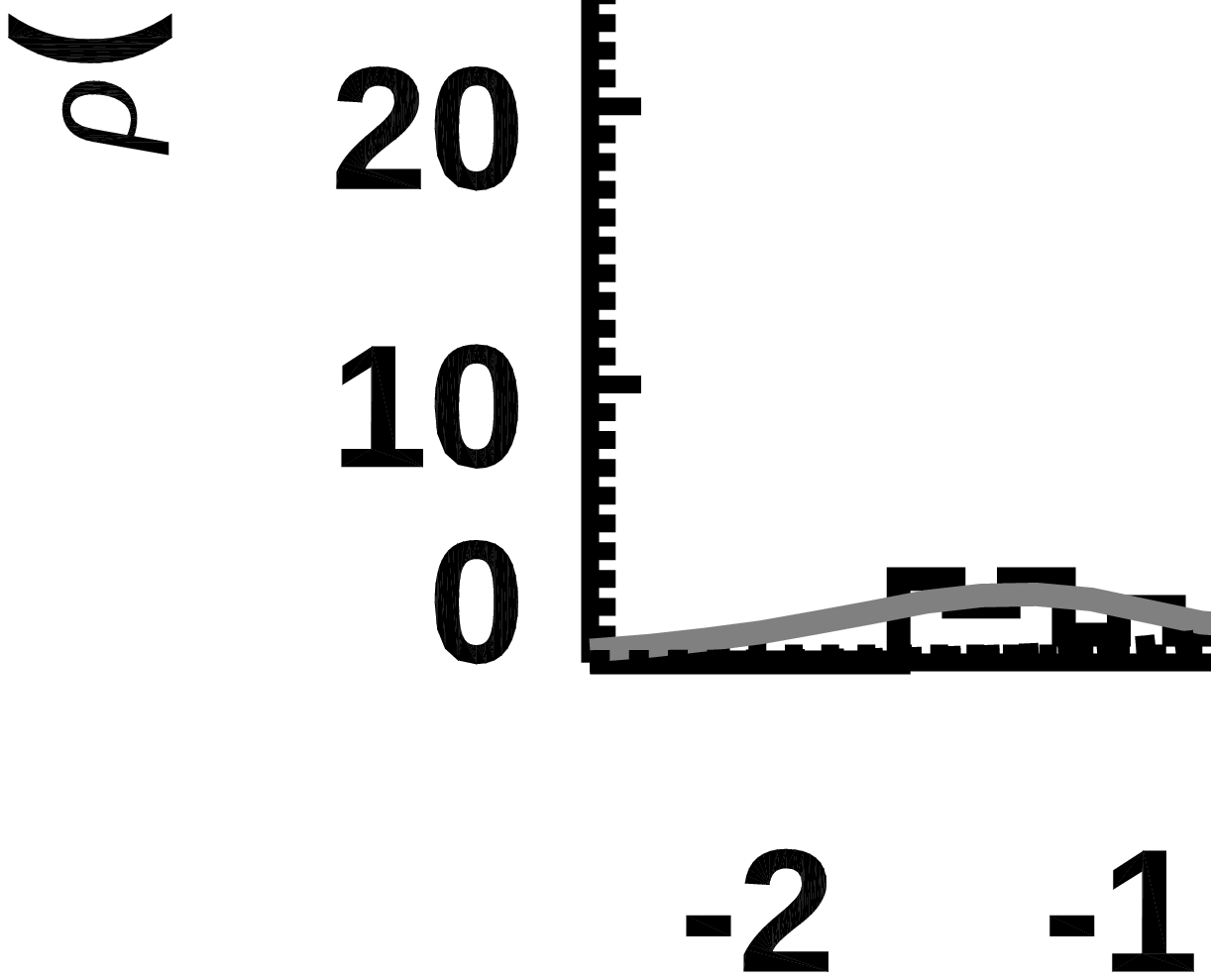} \\
\includegraphics[width=16 cm]{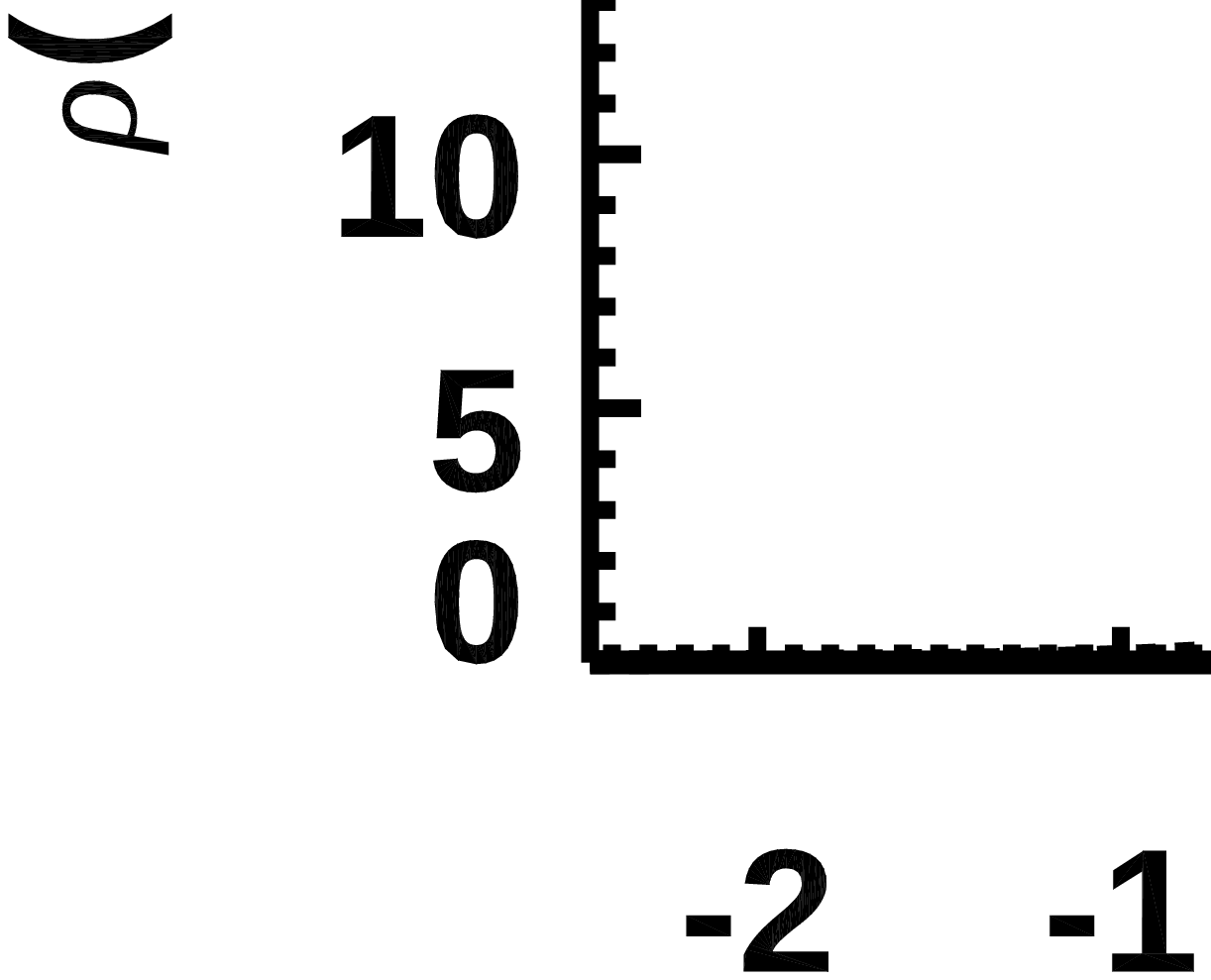} \\
\includegraphics[width=16 cm]{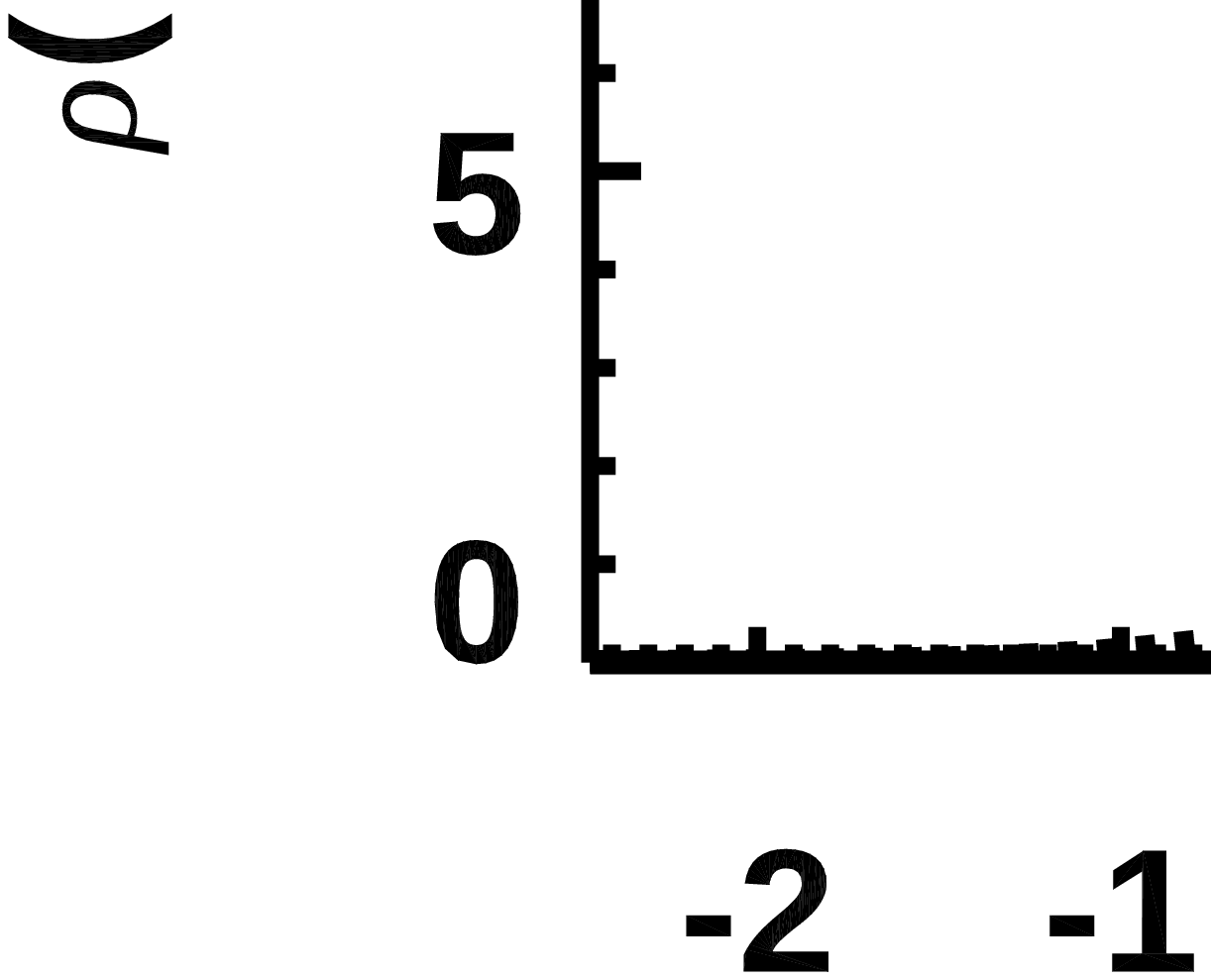} \\
\includegraphics[width=16 cm]{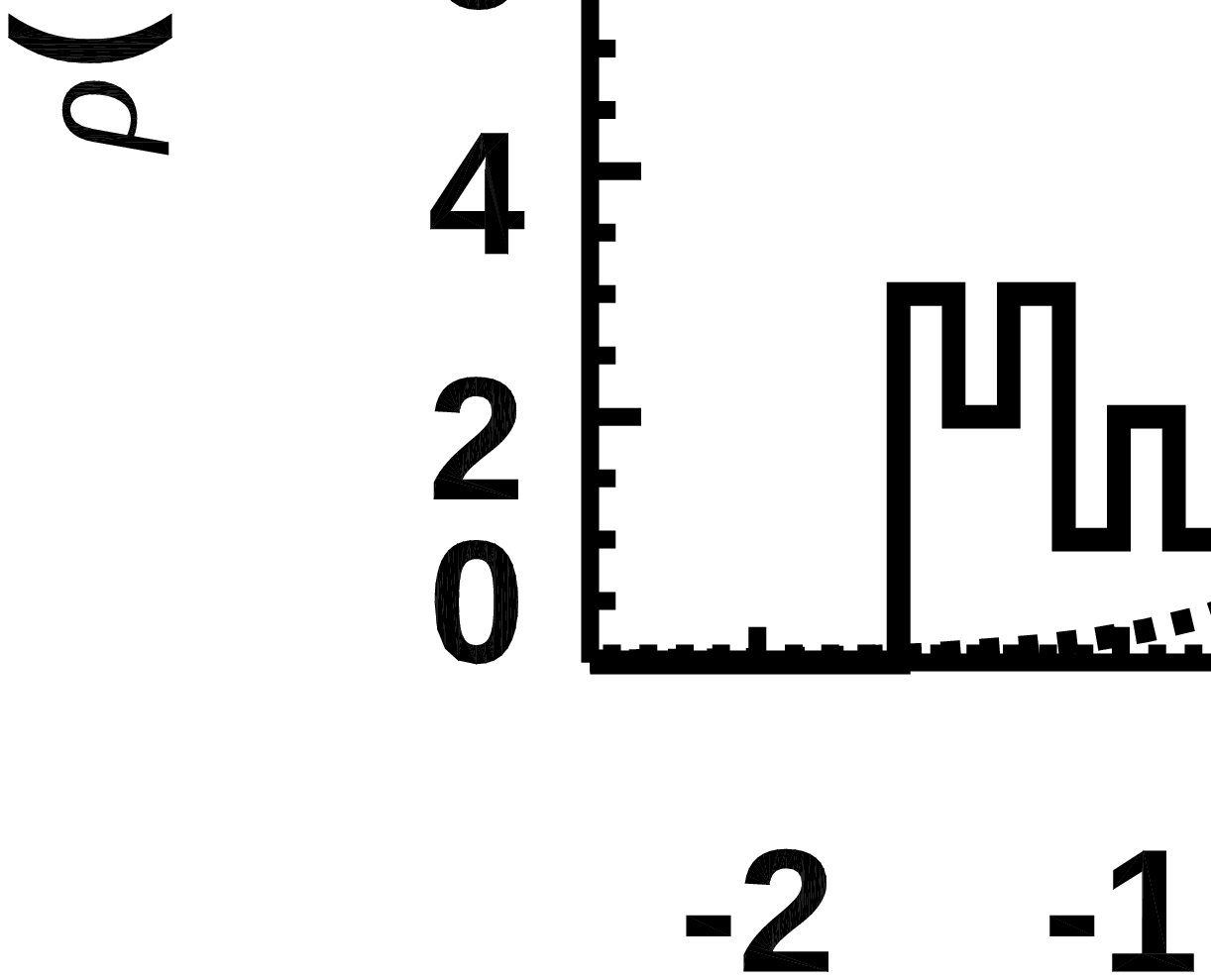}
\caption{Distribution of waiting times between gamma-ray flares of FSRQ detected with the \emph{iSRS} method (100 MeV sample):
Left columns report the distribution ($\rho(log(\Delta_t)$) of log($\Delta_t$). Right columns report the same distribution
divided by the binsize (this new distribution is proportional to
$\rho(log(\Delta_t))e^{-\Delta_t}$.
The top row reports the distribution for all the sources. The other rows report the distribution of waiting times for
sources selected on account of detected flares: from top to bottom $N_{flares} \le 9$,
$ 10 \le N_{flares} \le 19$, $N_{flares} \ge 20$.
Solid histograms represent the data, dashed curves the multi-loghat fitting function.
The solid grey line in the top row is the composite model obtained adding a poissonian process to the multi-loghat fitting function.
}
\label{fig_waiting_times}
\end{figure*}

In this paper we always report  waiting times, rates and duration in the host galaxy reference frame,
except where explicitly written.\\
We performed several uniformity and unimodality tests on the sets of activity peaks of the
apparently loudest sources (the sources with the largest number of detected flares).
They are reported in Table \ref{tab:kol}. In the assumption that the
FERMI-LAT regularly scanned each sky position, the distribution of flaring times cannot be
considered neither uniform, nor unimodal.\\
\begin{table*}
\caption{Tests for time distribution of flares on sources showing $N_{flares}\ >$ 15 in the 100 MeV sample:
Kolmogorov-Smirnov test of uniformity \citep[K-S][]{kolmogorov1933}, Frosini test of uniformity \citep{frosini1987}, Hartigan \& Hartigan test of unimodality \citep{hartigan1985}.
The test value (D$_{KS}$, B, D$_{HH}$) are reported for Kolmogorov-Smirnov, Frosini, and Hartigan \& Hartigan tests respectively.
The probability (p) is also reported for the null hypothesis of uniformity or unimodality.}
\label{tab:kol}
\begin{center}
\begin{tabular}{ l c l l l l l l l l}
source           &  \multicolumn{1}{l}{\#flares} & \multicolumn{2}{c}{K-S} & & \multicolumn{2}{c}{Frosini} & &\multicolumn{2}{c}{ Hartigan} \\ \cline{3-4} \cline{6-7} \cline{9-10}
                 &           & D$_{KS}$  & p &   &  B   & p & & D$_{HH}$      & p\\ 

PKS 1510-08      &   39      & 0.208 & 0.058               &  & 0.514 & 0.089                 & & 0.0756 & 0.070                 \\
3C 454.3         &   37      & 0.319 & 7.5$\times 10^{-4}$  &  & 0.634 & 0.031                 & &  0.082 & 0.045                \\
CTA 102          &   34      & 0.593 & 6.7$\times 10^{-6}$  &  & 1.80  & 2.2$\times 10^{-16}$  & & 0.0698  & 0.20                 \\
3C 279           &   28      & 0.634 & 9.0$\times 10^{-14}$ &  & 0.884 & 1.5$\times 10^{-3}$   & & 0.0997  & 0.020                \\
4C +21.35        &   18      & 0.379 & 7.7$\times 10^{-3}$  &  & 0.779 & 5.8$\times 10^{-3}$   & & 0.132  & 4.9$\times 10^{-3}$   \\
4C +71.07        &   17      & 0.327 & 0.040               &  & 0.614 & 0.035                 & & 0.131  & 7.9$\times 10^{-3}$    \\
\end{tabular}
\end{center}
\end{table*}
The obtained distribution of waiting times (with logarithmic bin) is shown
in Figure \ref{fig_waiting_times} (left column), while in the right column
is reported  the same distribution divided for the binsize;
the last distribution is proportional to $\rho(log(\Delta_t))e^{-\Delta_t}$.
Fitting functions are superimposed to the data.
The fitting function reported with dashed line represents
a set of overlapping bursts of flares (the multi-loghat distribution described in appendix A). The parameters
of the fitting functions are reported in Table \ref{tab:fit}.
The multi-loghat distribution is a poissonian process: the events
within a burst are uniformly distributed.
%
%
There is good agreement between data and the multi-loghat model for $\Delta_t$ larger than 1 day. For short waiting
times ($\Delta_t \rightarrow\ 0$) the multi-loghat distribution and all the distributions based on Poisson processes \citep[see, e.g.,][]{wheatlhand2000}
shows an exponential profile $\rho(\Delta_t)\rightarrow \frac{1}{\tau}e^{-\frac{\Delta_t}{\tau}}$
(where $\tau$ is the typical timescales of the distribution).
Thence, the fitting distribution reported in the right column of Figure \ref{fig_waiting_times} tends to a constant.
Experimental data show this trend (Figure \ref{fig_waiting_times}), but for
$\Delta_t\ <\ 1d$ data deviate from the typical Poissonian profile, and show an increasing trend for reducing values of $\Delta_t$.\\
We can regard the distribution
as combined of two distributions: a multi-loghat distribution and an other one acting on short waiting times
\citep[see, e.g., ][for the study of solar flares with ISEE-3/ICE]{aschwanden2010}.
We call fast component the second distribution; and we call multi log-hat + poissonian the composite distribution.\
Only a few tens of waiting times can be ascribed to the fast component,
thence we cannot establish a reliable functional form with this dataset. 
We simply added a further poissonian term responsible for short waiting times.
The resulting fitting function for the whole data sample is shown in Figure \ref{fig_waiting_times} (top row) and its parameters are reported in Table \ref{tab:fit}.
The observed fraction of short waiting times is reported as $R_{fast\ observed}$ 
in Table \ref{tab:fit}.
The reduction of the Cash estimator obtained adding the fast component
tells us that the fast component is statistically relevant.\\
We tested also other fitting functions for the temporal distribution of flares. We obtained an intermediate result
in term of Cash estimator with the multi-pow distribution function (defined in appendix A). Results are reported in Table \ref{tab:fit}.
The multi-pow distribution, indeed, does not reproduce the trend observed for $\Delta_t\ < 1d$. It gives better
results with respect to the \emph{multi log-hat} distribution, because for the \emph{multi-pow} distribution,
the function $\rho(log(\Delta_t))e^{-\Delta_t}$ rises while $\Delta_t$
decreases. Thence, it roughly reproduces data with short $\Delta_t$. \\
We also tried to fit with another two-components distribution: the multi-pow + poissonian (defined in appendix A, with 5 parameters).
Results are reported in Table \ref{tab:fit}.
With this distribution we obtained the best Cash-estimator, but it is statistically comparable with the value obtained with the multi-loghat + poissonian.
The multi-loghat+poissonian and the multi-pow distributions (with 5 parameters)
are the statistically relevant models: in fact, they
give $\Delta C=-10.0$, and  $\Delta C=-11.5$ respectively with respect to the multi-pow distribution (with 3 parameters).\\
\begin{figure*}
\includegraphics[width=8 cm]{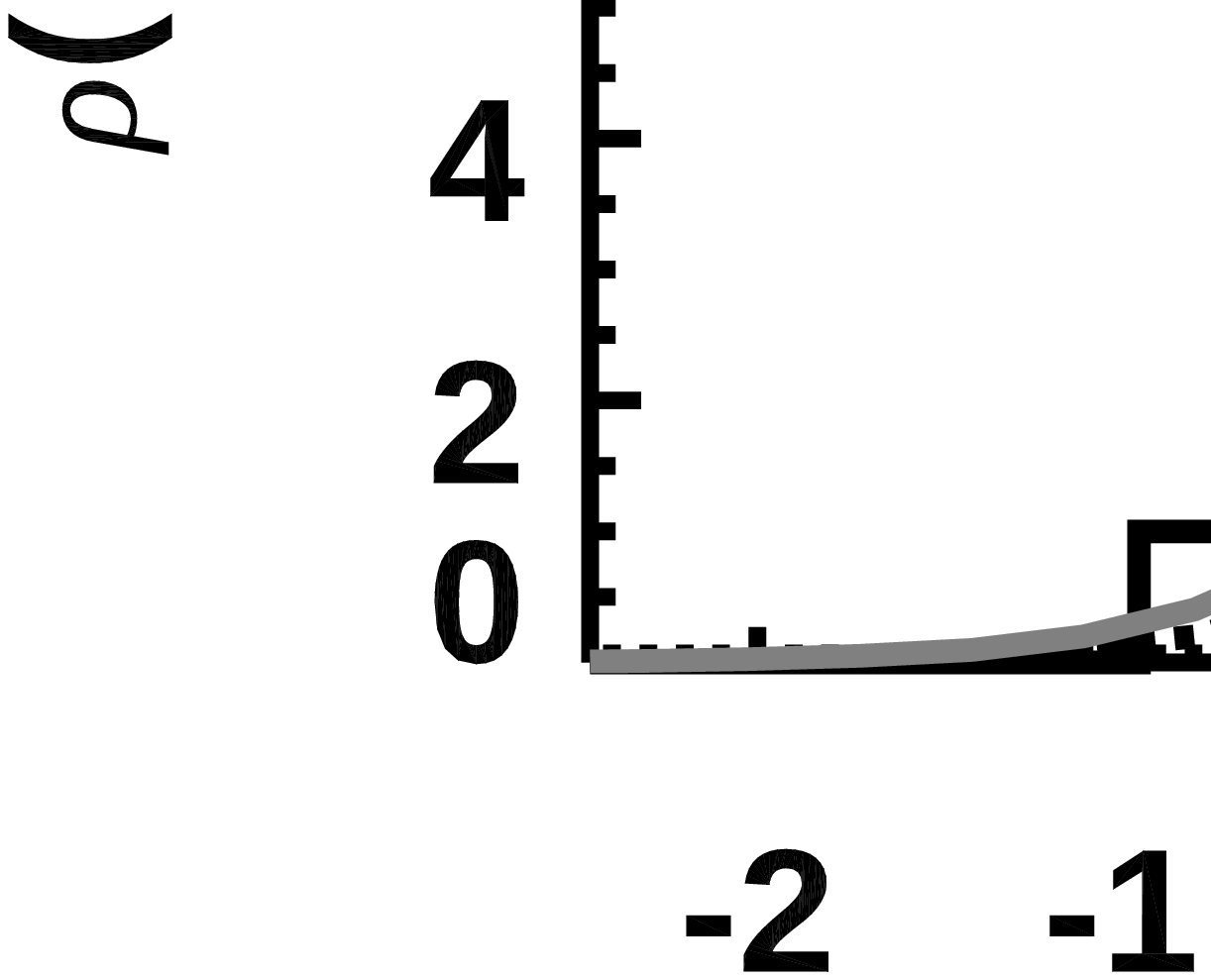} \includegraphics[width=8 cm]{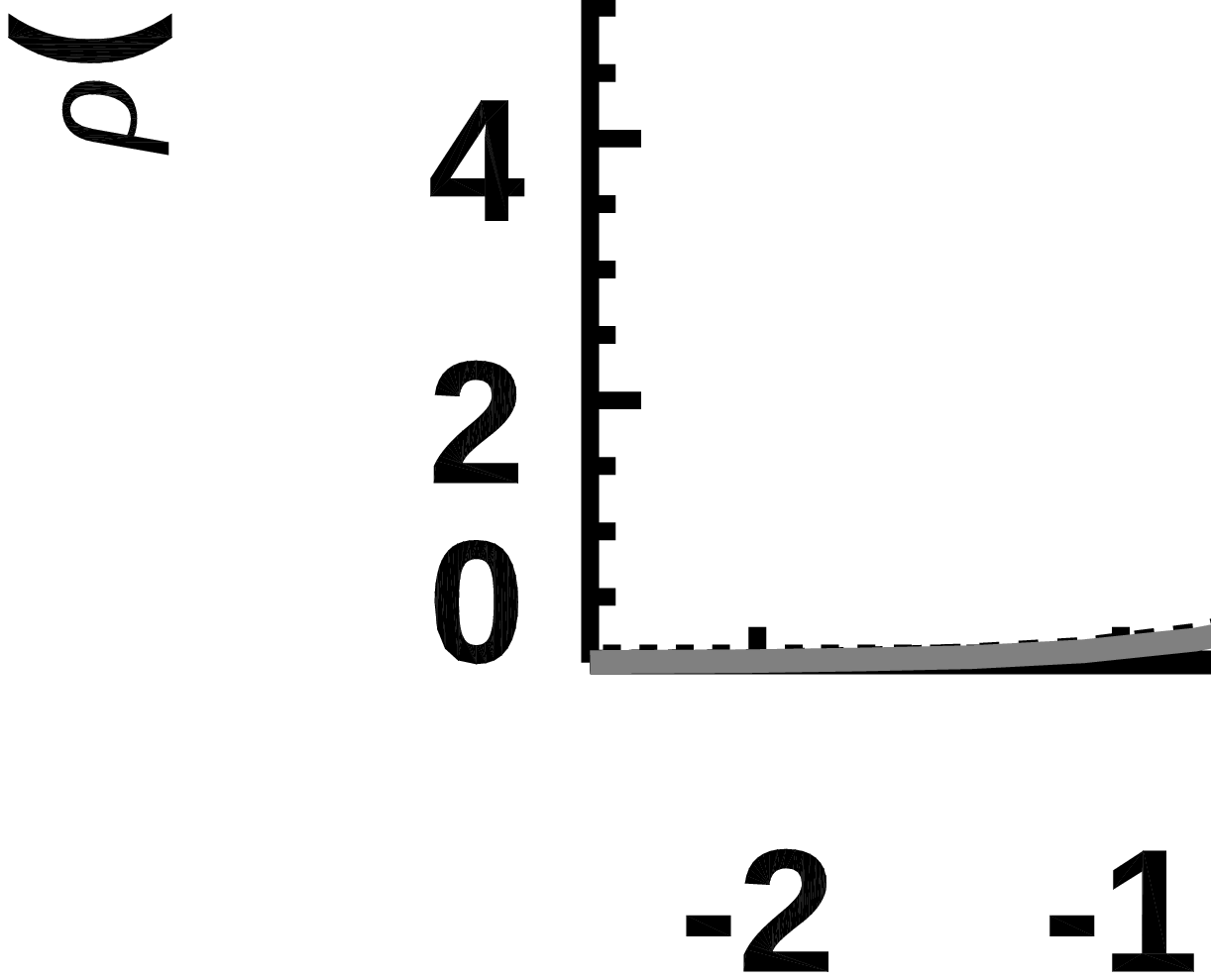} \\
\includegraphics[width=8 cm]{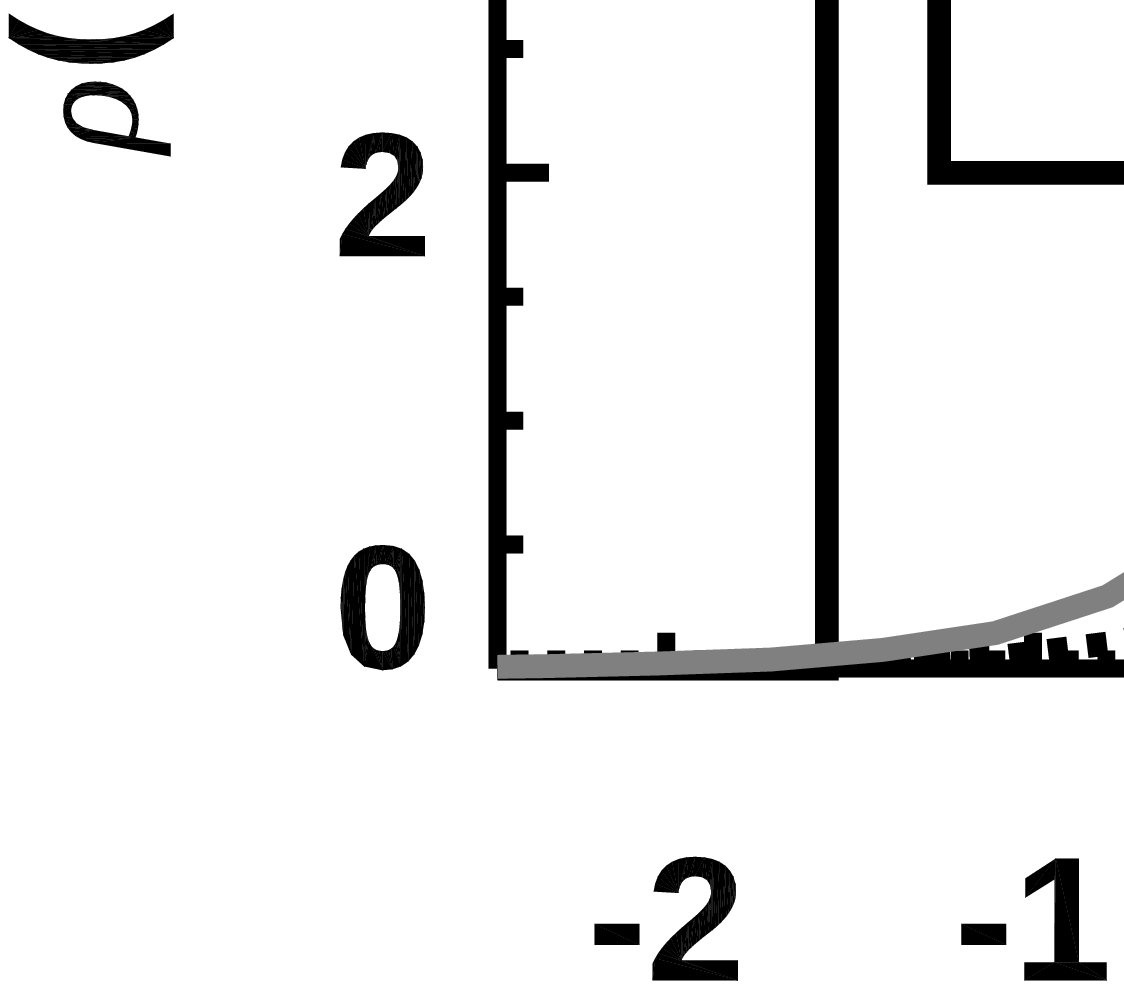} \includegraphics[width=8 cm]{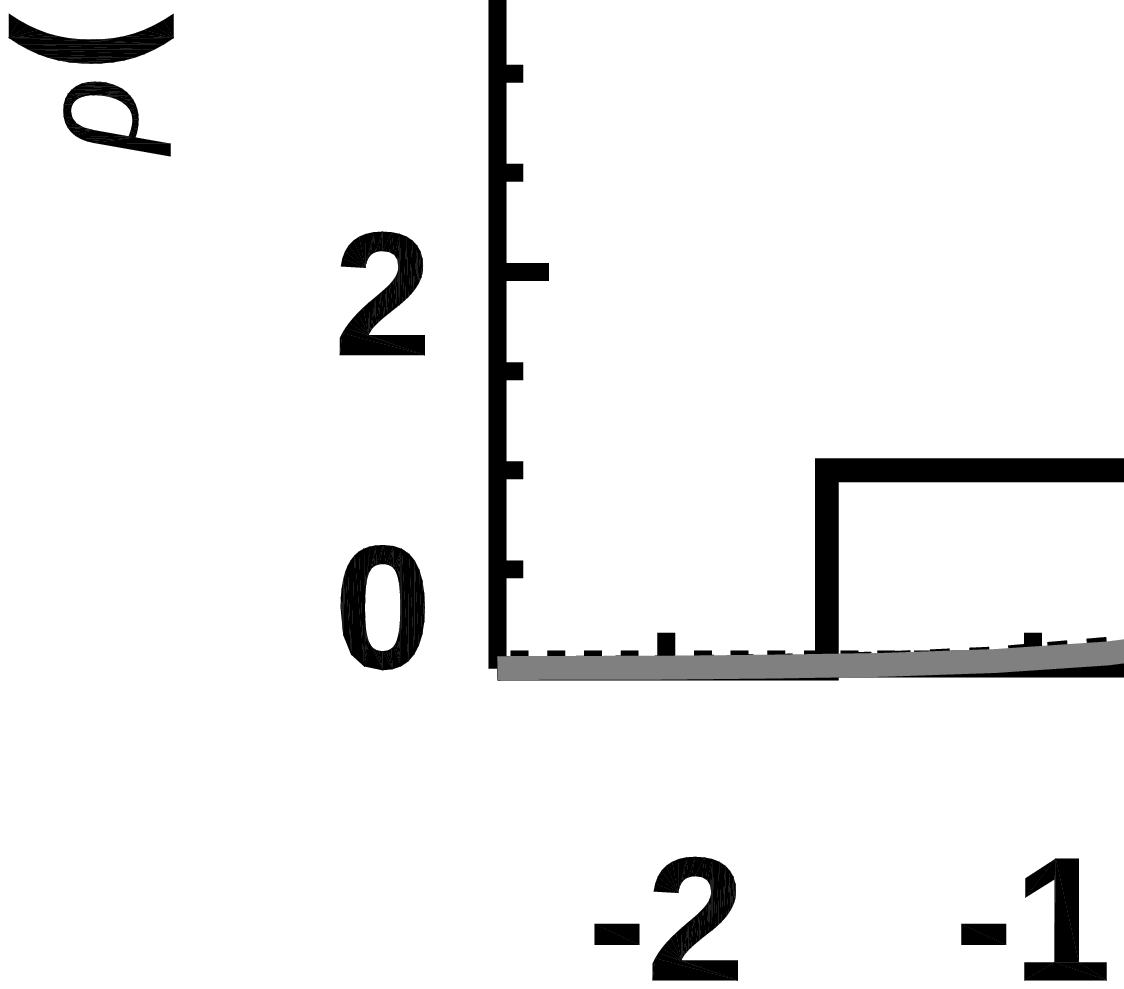} \\
\includegraphics[width=8 cm]{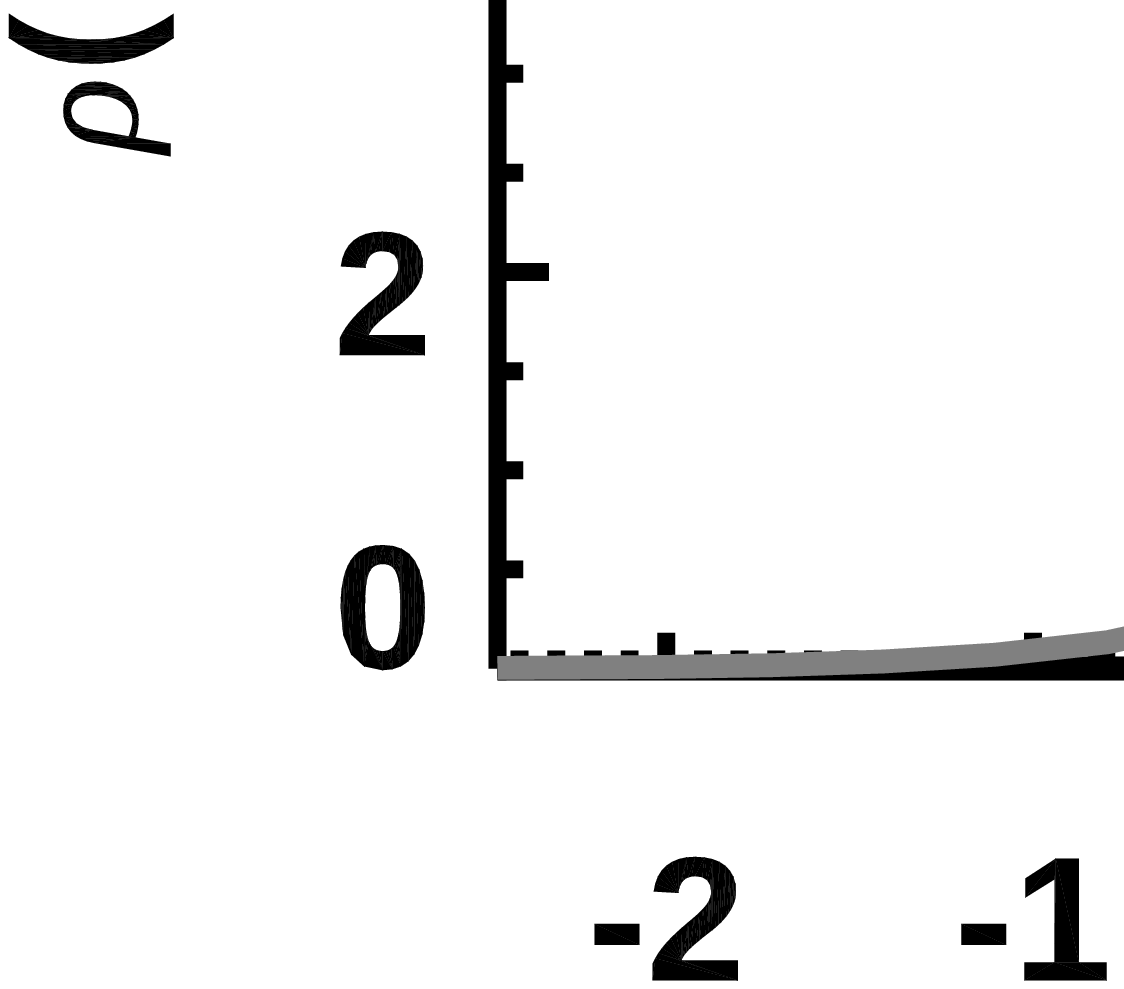} \includegraphics[width=8 cm]{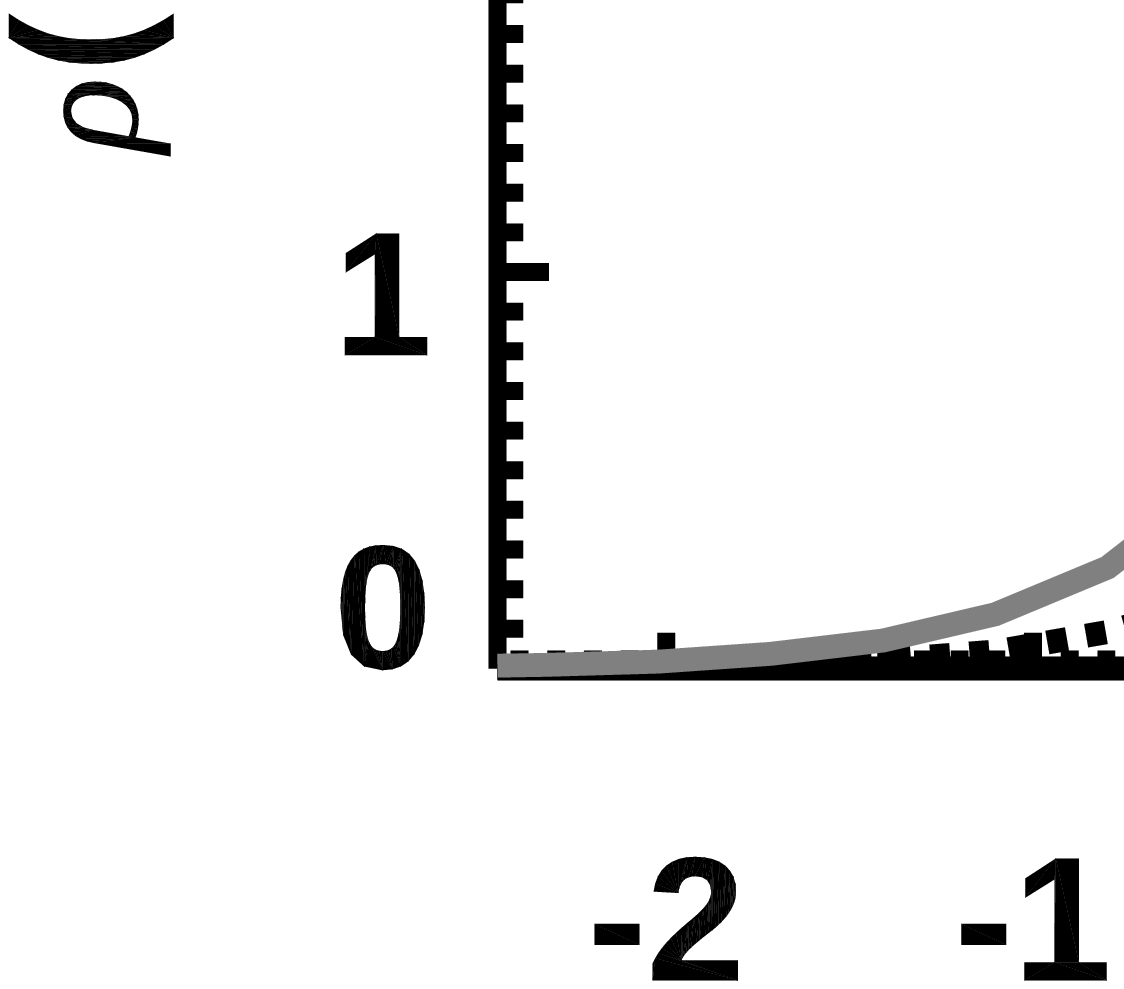} \\
\includegraphics[width=8 cm]{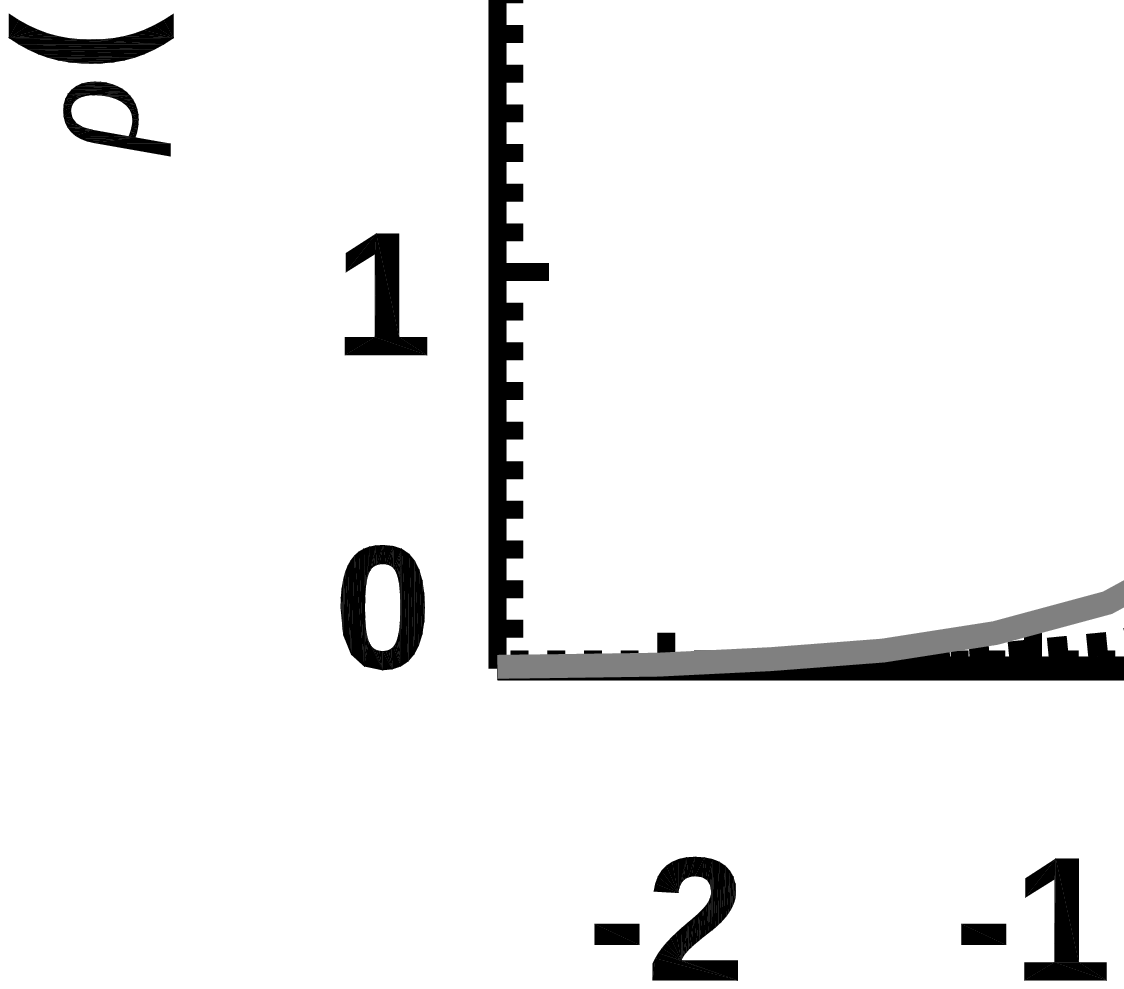}  \includegraphics[width=8 cm]{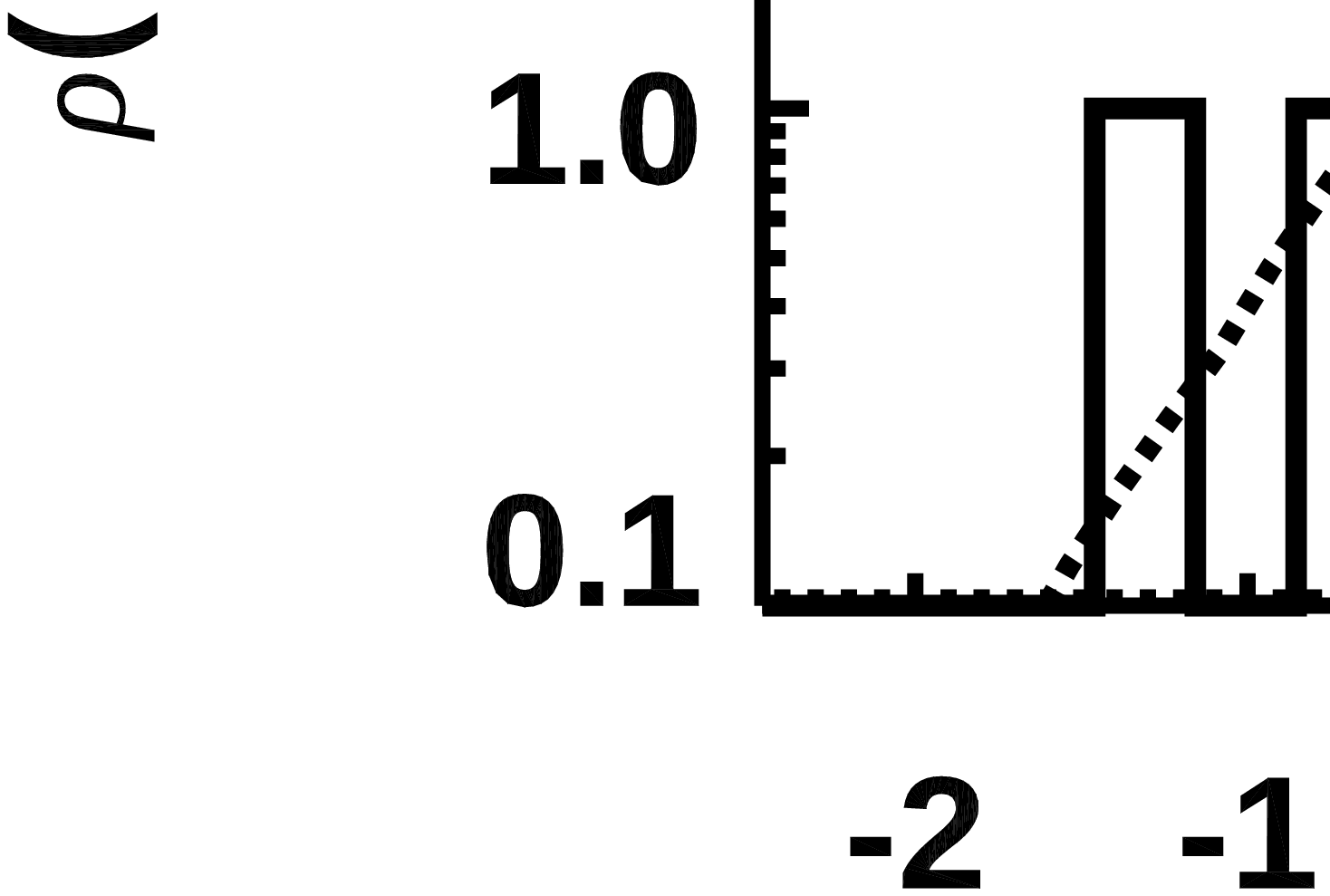} \\\\
\caption{Distribution of waiting times for the apparently loudest sources in the sample, and for the whole sample (excluding CTA 102, right bottom plot). 
Solid histograms represent the data, dashed curves the multi-loghat fitting function evaluated for the whole data sample. For single sources, the grey lines represent the multi-loghat fitting function evaluated for each source separately, with parameters reported in Table \ref{tab:fit_single}; fit is performed for $\Delta_t>0.3$d to reduce the contribution of the fast component.
}
\label{fig_waiting_times_single}
\end{figure*}
%
%
%
We have shown in appendix C that the resolving power depends on the temporal distance between flares, and on peak fluxes.
Moreover, the apparently loudest sources are also the brightest ones. Thence we expect that the apparently loudest
sources show the largest amount of short waiting times.
This trend shows up in Figure \ref{fig_waiting_times}.\\
We will show for a physical scenario that close flares could not be resolved individually (pile-up effect), causing a reduction of short waiting time statistics.\\
\subsection{a case study: CTA 102}
\begin{figure}
\includegraphics[width=8 cm]{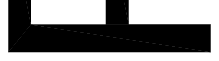} \\
\includegraphics[width=8 cm]{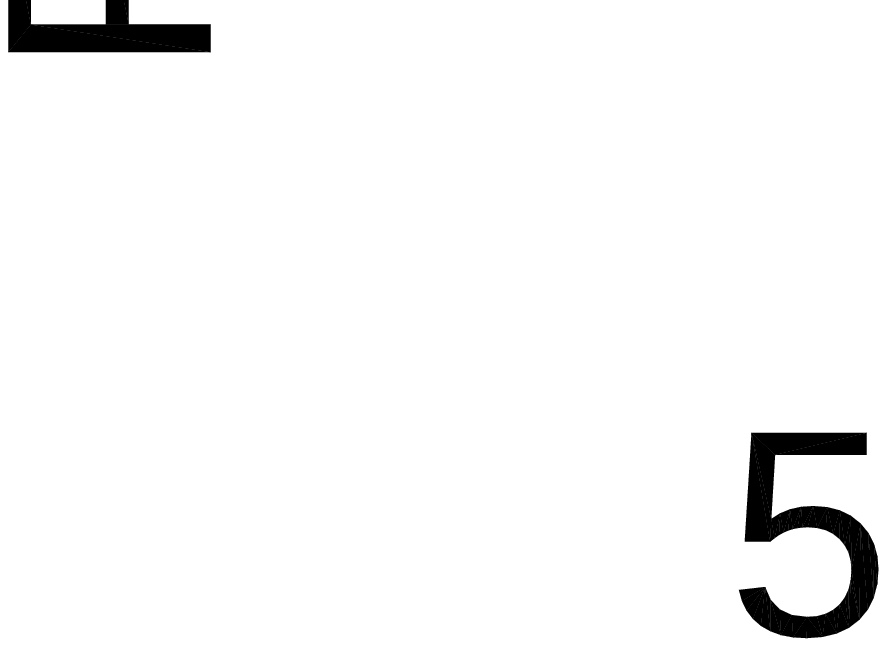} \\
\includegraphics[width=8 cm]{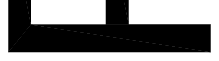} \\
\caption{Unbinned light curve \citep{pacciani2018} for CTA 102 obtained for E$>$100 MeV \citep[see][ for a detailed explanation]{pacciani2018}, and zooms of relevant periods.
The horizontal segments represent statistically relevant clusters of events obtained with the \emph{iSRS} procedure.
They are described by  duration, average photometric flux, and median time (or by starting time). Vertical lines are the error bars on flux estimate.
The set of clusters form a tree called \emph{unbinned light curve}. Background is not subtracted}
\label{fig_ulc_cta102}
\end{figure}

We note that the majority of short waiting times ($\Delta_t\ <\ 1d$) come from $CTA 102$. They are mainly grouped around MJD 57738 and MJD 57752.
The unbinned light curve of the source for a 100d period (obtained with a chance probability threshold $P_{thr}\ =\ 1.3\times10^{-3}$) is reported in fig \ref{fig_ulc_cta102}.
The peculiar waiting times distribution is reported in Figure \ref{fig_waiting_times_single}.\\
The unbinned light curve can be compared with the results reported in \citealt{dammando2019} and in \citealt{meyer2019}. There are 2 fast flares
that were fully resolved with the \emph{iSRC}:
the first flare peaks at MJD 57752.50 with a duration of 7.8 hours (FWHM, observer frame), 
the second one peaks at MJD 57752.83 with a duration of 1.5 hours (FWHM, observer frame).
A further fast flare is fully resolved adopting
a $P_{thr}\ =\ 2.3\%$: peaking at MJD 57738.00, with a duration of 2 hours. We further note that in the 0.3d period between MJD 57738.2 and 
MJD 57738.5 the \emph{iSRS} procedure resolves a set of 4 peaks (but the statistics does not allow to fully resolve flares). Thence,
the average duration of each flare of the set is $\le\ $0.1d. Similarly, in the 1d period between 57751.8 and 57753.0 there are 3 resolved and 2 fully
resolved flares, for an average flare duration of $\le$0.2d.\\ 
We discuss in appendix C the temporal pile-up. In particular flares with a temporal distance of 0.1d could be resolved
if their flux exceed 2$\times10^{-5}\ ph\ cm^{-2}\ s^{-1}$ (Figure \ref{fig_pileup}). This is the case for the resolved flares just mentioned. We cannot exclude that
the tail preceding the peak at MJD 57738.00 is due to the pile-up of several flares with a trend of decreasing waiting times, or a trend of increasing flux.
In fact, there is an hint of a complex structure from the 12h binned light curve reported in \citealt{dammando2019}.\\
\subsection{a peculiar source: S4 0218+35}
\begin{figure}
\includegraphics[width=8 cm]{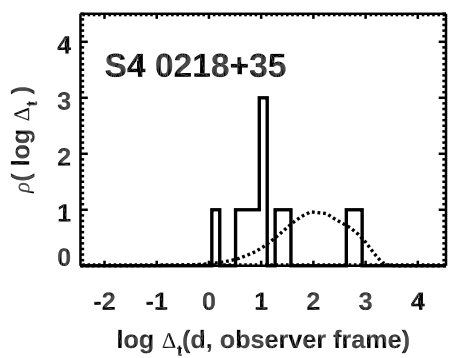}
\caption{Distribution of waiting times for the lensed source  S4 0218+35 (observer reference system).
Solid histogram represents the data, dashed curve the fitting function 
}
\label{fig_waiting_times_S4_0218p35}
\end{figure}

The sample of sources we analysed includes also S4 0218+35, a blazar with a gravitational lens along the line of sight. The effect of the gravitational lens shows up in the distribution of waiting times reported
in Figure \ref{fig_waiting_times_S4_0218p35}.
They cluster around $\Delta_t \ \sim11$d (observer frame), in close agreement with the
waiting time of delayed echo measured with VLA \citep{biggs1999,cohen2000} and with the gamma-ray data itself \citep{barnacka2016}.\\

\subsection{remaining sources}
\begin{figure}
\includegraphics[width=8 cm]{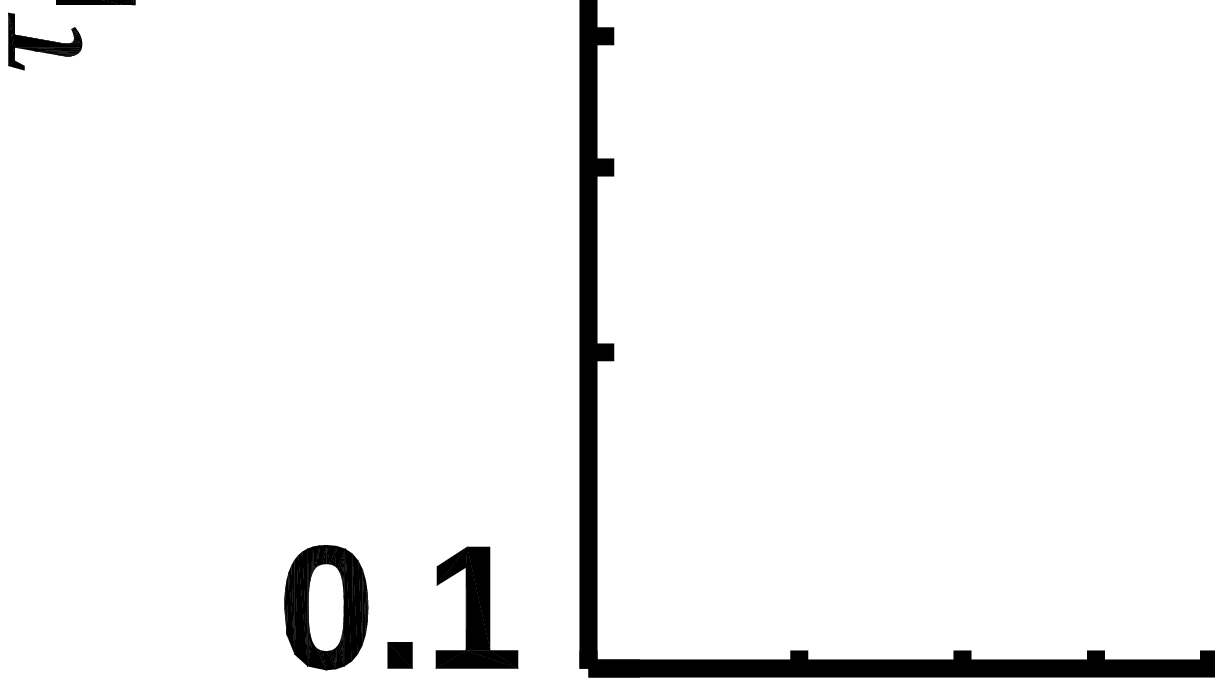} \\
\includegraphics[width=8 cm]{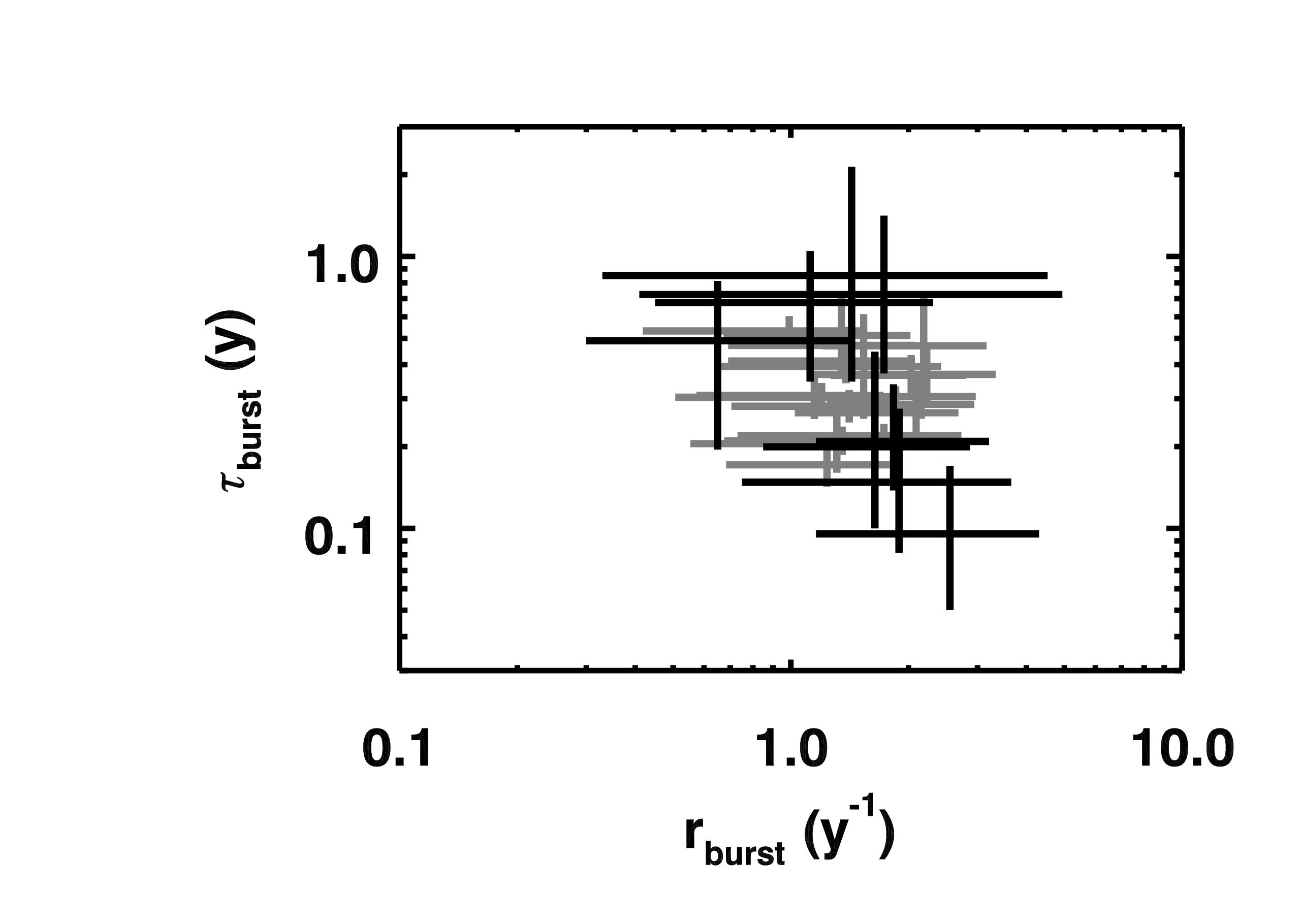} \\
\caption{Top panel:
  multi-loghat fit of waiting times for the 100 MeV sample:
  joint confidence region for $r_{burst}$ and $\tau_{burst}$. 
  The parameters for the minimum of the Cash estimator are reported with a cross.
  Contours are reported for 90\% and 99\% confidence levels. 
  The resolved superluminal features ejection rate and fading time observed at 43 GHz \citep{jorstad2017} are reported with
  a diamond symbol (average values). Bottom panel:
  $r_{burst}$ and $\tau_{burst}$ (represented with black data)
  obtained by fitting the multi-loghat distribution to the data of single sources (for the apparently loudest sources in
  our gamma-ray sample). For comparison, the apparent ejection rate and fading times for each FSRQs reported in \citep{jorstad2017}
  are reported too (grey data).
}
\label{fig_ulen_vs_rate}
\end{figure}

In order to reduce the signal from the short component, we performed a further investigation of waiting times distribution
removing CTA 102 from the sample. \\
We evaluated several fitting functions. The relevant are described in appendix A.
In Figure \ref{fig_waiting_times} we show the fit with the multi-loghat distribution. 
It gives a satisfactory description of data;
results are reported in Table \ref{tab:fit};
\begin{table*}
\caption{Fitted values of the parameters of the multi-loghat, multi-loghat+poissonian, multi-pow, multi-pow+poissonian distributions for the temporal distribution of flares (reported statistical errors with 90\% confidence level, systematic errors are reported when they comparable to statistical errors).
$C$ is the value of the Cash estimator at minimum.
For multi-loghat model:
$r_{burst}$ is the burst rate in $y^{-1}$ , $m_{log(\tau)}$ and $\sigma_{log(\tau)}$ are the mean and standard deviation of the gaussian distribution for $log(\tau_{burst})$, with $\tau_{burst}$ in $y$.
For multi-loghat + poissonian: the additional parameters are the fraction of events  due to the fast component ($R_{fast\ observed}$),
and the typical timescale $\tau_{fast}$  of the fast component in hours , modelled with a poissonian process.
For multi-pow model:
$r_{burst}$ is the burst rate in $y^{-1}$, $\tau_{burst}$ is the burst duration in $y$, $\eta$ coefficient is described in appendix A.
For multi-pow + poissonian: the additional parameters are the fraction of events  due to the fast component ($R_{fast\ observed}$),
and the typical timescale $\tau_{fast}$  of the fast component in hours, modelled with a poissonian process.
}
\label{tab:fit}
\begin{center}
\begin{tabular}{l l l l l l}
\multicolumn{4}{c}{multi-loghat (all sources)}  \\  \cline{1-4} \\
$C$          & $r_{burst}$ & $m_{log(\tau)}$ & $\sigma_{log(\tau)}$ \\
             & $(y^{-1})$ &  &  \\
-2746.1      &
1.3$\pm$0.4(stat)  & -0.21$_{-0.13}^{+0.18}$(stat) & 0.36$\pm$0.14(stat) \\
\\ \\
 \multicolumn{4}{c}{multi-pow (all sources)}  \\  \cline{1-4} \\
 $C$          & $r_{burst}$ &$\eta$ & $\tau_{burst}$ \\ 
             & $(y^{-1})$  &        & $(y^{-1})$ \\ 
 -2766.0     &
1.5$\pm$0.3 (stat) & 3.0$_{-0.2}^{+0.5}$(sys)$_{-0.5}^{+0.4}$(stat) &  0.67$_{-0.00}^{+0.13}$(sys)$_{-0.16}^{+0.24}$(stat)  \\
\\ \\
\multicolumn{4}{c}{multi-loghat + poissonian (all sources)}  \\  \cline{1-6} \\
$C$          & $r_{burst}$ & $m_{log(\tau)}$ & $\sigma_{log(\tau)}$ & R$_{fast\ observed}$ & $\tau_{fast}$ \\
             & $(y^{-1})$ &               &                    & $(\%)$ & $(hr)$\\
-2776.0      & 1.3$\pm$0.3(stat)  & -0.20$_{-0.12}^{+0.08}$(stat) & 0.35$\pm$0.13(stat) & 2.9$_{-0.11}^{+0.10}$(stat) & $1.2_{-0.3}^{+0.1}$(sys)$_{-0.5}^{+1.6}$(stat) \\
\\ \\
 \multicolumn{4}{c}{multi-pow + poissonian (all sources)}  \\  \cline{1-6} \\
 $C$          & $r_{burst}$ &$\eta$ & $\tau_{burst}$  & R$_{fast\ observed}$ & $\tau_{fast}$ \\ 
              & $(y^{-1})$ &   & $(y^{-1})$        & $(\%)$              & (hr)\\ 
 -2777.5     &
1.4$\pm$0.3 (stat) & 2.1$_{-0.0}^{+0.4}$(sys)${\pm 0.5}$(stat) &  0.65$_{-0.00}^{+0.09}$(sys)$_{-0.15}^{+0.22}$(stat)   & 2.5$_{-0.0}^{+0.6}$(sys)$_{-1.0}^{+1.8}$(stat) & $0.8_{-0.1}^{+0.2}$(sys)$_{-0.4}^{+1.3}$(stat) \\
\\ \\ \\
\multicolumn{4}{c}{multi-loghat (CTA 102 excluded)}  \\  \cline{1-4} \\
$C$          & $r_{burst}$ & $m_{log(\tau)}$ & $\sigma_{log(\tau)}$ \\
             & $(y^{-1})$ &               &  \\
-2642.1     &
1.3$\pm$0.4(stat)  & -0.21$_{-0.13}^{+0.18}$(stat) & 0.36$\pm$0.14(stat) \\
\\ \\
 \multicolumn{4}{c}{multi-pow (CTA 102 excluded)}  \\  \cline{1-4} \\
 $C$          & $r_{burst}$ &$\eta$ & $\tau_{burst}$ \\ 
              & $(y^{-1})$ &        & $(y^{-1})$ \\ 
 -2643.8     &
1.4$\pm$0.3(stat) & 1.9$_{-0.0}^{+0.4}$(sys)$_{-0.4}^{+0.6}$(stat) &  0.64$_{-0.00}^{+0.18}$(sys)$_{-0.10}^{+0.26}$(stat) \\ \\
\end{tabular}
\end{center}
\end{table*}
the joint confidence region for burst rate and duration parameters is reported in Figure \ref{fig_ulen_vs_rate} for the multi-loghat distribution (top panel). Fitting  the multi-loghat function to the waiting time distribution  generates correlated estimates of the two parameters.\\
The Cash estimator variation from an uniform to a multi-loghat function is $\Delta C$=-105.8, in favour of the multi-loghat distribution.\\
We tested also other fitting functions for the temporal distribution of flares. We obtained slightly better results with the multi-pow
distribution function. Results are reported in Table \ref{tab:fit}.
Furthermore, we obtained that if fitting is limited to $\Delta_t>$1d, Cash estimators for multi-loghat and multi-pow distribution become very similar.\\
Changing the pole position into the middle of the burst for the multi-pow does not produce a better value of the Cash estimator.
Moreover, using  multi-pow-smooth instead of a multi-pow does not give better Cash estimator.
In Figure \ref{fig_waiting_times_single} we show the waiting times distribution of the apparently loudest sources in our sample, together with the multi-loghat
fitting function, evaluated at minimum of the Cash estimator for the whole 100 MeV sample, except CTA 102.
A small amount of short waiting times are well outside the fitting distribution of single sources: the fast component arises also for sources other than CTA 102, but
close in time activity peaks are resolved for extremely bright flares only. This is the case for 3C 454.3 flaring around MJD 55517-55521 and for 3C 279,
flaring around MJD 57189.\\
The fast component does not show up in the distribution of waiting times accumulated for all the sources (except CTA 102).
We will show that for realistic simulations (taking into account flare duration, and exposure variation with time), some pileup effect appears for $\Delta_t$ shorter than several days.
The net effect is a reduction of statistics for short waiting times. We speculate that the cumulative distribution reported in Figure \ref{fig_waiting_times_single}
(right bottom plot) does not show the pile-up effect, because the fast component compensates for the reduction of statistics for short waiting times.\\
In Table \ref{tab:fit_single}, in Figure \ref{fig_waiting_times_single} and \ref{fig_ulen_vs_rate} (bottom panel) we report the results obtained fitting the multi-loghat model to the waiting time distributions for single sources. Due to the limited sample size,
we fitted model to the data of single sources fixing the parameter $\sigma_{log(\tau)}$ to the value obtained by fitting the multi-loghat model to the accumulated distribution ($\sigma_{log(\tau)}=0.36$). 
Moreover, we restricted to $\Delta_t\ge$0.3d to minimize the contribution of the fast component to the single source samples.\\
\begin{table}
\caption{Fitting parameters of the multi-loghat model for the temporal distribution of flares (reported statistical
errors with 68\% confidence level). The first column reports the source name for which the fit was performed. The second column 
reports the value of the obtained minimum for the cash estimator.
$r_{burst}$ is the burst rate in $y^{-1}$, $m_{log(\tau)}$ is the mean of the gaussian distribution
for $log(\tau_{burst})$, with $\tau_{burst}$ in y. The standard deviation of the distribution of $log(\tau_{burst})$ is held fixed
at the value obtained fitting the multi-loghat function to the accumulated distribution of waiting times
($\sigma_{log(\tau)}=0.36$). To keep at a minimum the contribution of the short waiting time component, fit is performed
on the reduced sample with $\Delta_t > 0.3d$.}
\label{tab:fit_single}
\begin{center}
\begin{tabular}{lrll}
source         &   cash   & $r_{burst}$             & $m_{log(\tau)}$ \\ \hline
               &          &  $(y^{-1})$             &  \\ 
PKS 1510-08    &   -43.3  &   1.8$_{-0.7}^{+1.4}$    & -0.68$_{-0.18}^{+0.21}$ \\\\
3C 454.3       &   -22.2  &   1.1$_{-0.7}^{+1.2}$    & -0.17$_{-0.29}^{+0.19}$ \\\\
CTA 102        &     5.4  &   2.5$_{-1.4}^{+1.8}$    & -1.02$_{-0.28}^{+0.25}$ \\\\
3C 279         &    -3.2  &   1.7$_{-1.3}^{+3.2}$    & -0.14$_{-0.29}^{+0.29}$ \\\\
4C +21.35      &    -0.5  &   0.65$_{-0.35}^{+0.78}$ & -0.31$_{-0.40}^{+0.22}$ \\\\
4C +71.07      &    10.1  &   1.9$_{-1.1}^{+1.8}$   & -0.83$_{-0.26}^{+0.27}$ \\\\
4C +38.41      &    11.0  &   1.6$_{-0.8}^{+1.2}$   & -0.70$_{-0.30}^{+0.35}$ \\\\
PKS 0402-362   &    11.0  &   1.4$_{-1.1}^{+3.1}$   & -0.07$_{-0.39}^{+0.40}$ \\
\end{tabular}
\end{center}
\end{table}
\section{Discussion and Conclusion}
\subsection{Waiting Times distribution for a non-ideal instrument}
\begin{figure}
\includegraphics[width=8 cm]{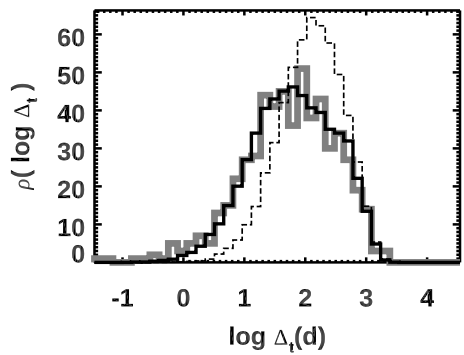}
\caption{
  Distribution of waiting times for data (grey line), the simulated case \emph{a} (uniformly distributed flares, thin dashed line), and simulated case \emph{c}$_{lct}$
  (light crossing time prevails, black continuous line).
}
\label{fig_data_vs_logstep_vs_unif}
\end{figure}

\begin{figure}
\includegraphics[width=8 cm]{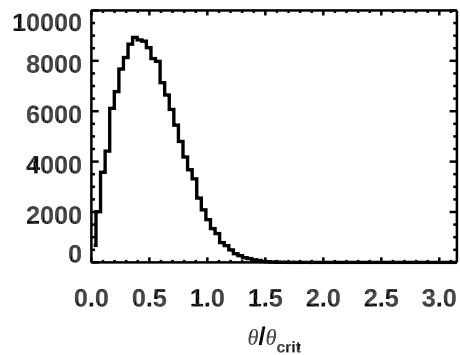}\\
\includegraphics[width=8 cm]{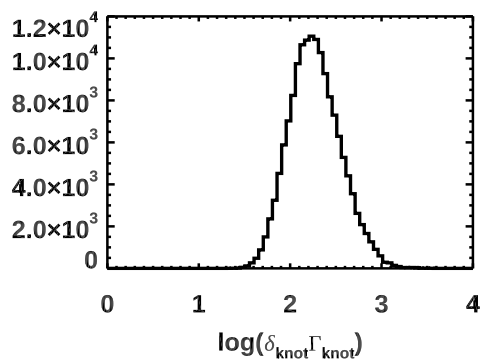} 
\caption{Top panel: Distribution of viewing angles of gamma-ray detected flares obtained simulating case {\emph c}$_{cooling}$.
Bottom panel: Distribution of $\Gamma_{knot}\delta_{knot}$ for gamma-ray detected flares obtained simulating case
{\emph c}$_{cooling}$.
To produce the histogram, the observer line of sight with respect to the jet axis was generated uniformly within a
cone of aperture 30$^\circ$.
}
\label{fig_knot_viewing_simulated}
\end{figure}

In the previous section we studied the waiting time distribution for an ideal instrument.
In order to investigate the effect of the exposure variation with time for
each source, the gaps between observations
of the FERMI-LAT, and the pile-up effect,
we performed detailed simulations taking into account for the exposure to each source.
We investigated three cases:\\
\emph{ a)} Flares uniformly distributed within the observation period of 9.5 y.\\
\emph{ b)} Time distribution of flares follows the multi-loghat distribution.\\
For both case \emph{a} and \emph{b}
we extracted the logarithm of the Doppler factor of accelerated electrons $log(\delta_{flare})$ with a gamma distribution, with mean=1, and variance=0.045 (see the best-fit intrinsic jet properties obtained for the 1.5JyQC quasar sample in \citealt{lister2019}).
Flares duration is proportional to $\delta_{flare}^{-1}$.\\
The observer line of sight with respect to the jet axis, and the bulk Lorentz factor of the emitting feature
affect the observed peak flux, flare duration, and
waiting times between flares. The amount of the effect depends on the emission process responsible for gamma-ray emission,
and on the physical scenario responsible for flaring activity. To account for this, we also studied a simple geometry for the
emitting source.
(case {\emph c} below):\\
{\emph c} Time distribution of flares follows the multi-loghat distribution:
Flares are uniformly generated within the burst.
Flares within each burst of activity have all the same
 Lorentz factors, and the same orientation (of the moving blob) with respect to the jet axis: it is uniformly extracted
within a cone of aperture 1$^{\circ}$ around the jet axis.
The line of sight is uniformly extracted within a cone of 8$^{\circ}$ aperture  from the  axis of the blazar jet.
For each flare,  peak luminosity
and flare duration FWHM are anticorrelated. We explored two  sub-cases: peak luminosity is proportional to $\delta_{flare}^{4+2\alpha}$ 
\citep[EC leptonic model, see, e.g.][]{dermer1997}, and the FWHM is proportional to $\delta_{flare}^{-1}$
(light crossing time prevails on cooling, case \emph{c}$_{lct}$);
or the flare duration FWHM is proportional to $\Gamma_{flare}^{-\frac{3}{2}}\delta_{flare}^{-\frac{1}{2}}$
\citep[cooling time prevails, case \emph{c}$_{cooling}$, see, e.g.,][]{pacciani2014}.\\
The burst rate is a constant for all the sources.
We didn't try to reproduce the fast component of waiting times, but only the multi-loghat component.
This last case imposes that all the flares from the same burst have the same Doppler factor,
and thence have a large coherence.\\
We modelled peak luminosity and flare duration of simulations with a parametric distribution, whose
parameters were obtained by fitting the models to the peak-luminosity and duration of detected flares.
Further details of the models will be described in a dedicated paper.
For all the models, flare shape was modelled according to Equation 7 in \citealp{abdo2010},
and converted in a temporal distribution of gamma-rays
detected by an ideal telescope. We tried also to model the flare temporal profile with gaussian and gamma distributions,
obtaining similar results.
Finally each gamma-ray is randomly accepted or rejected according to the real exposure to the source.
Exposure was evaluated for temporal bins of 86.4s.
Once the time series were obtained for each source,
we applied the {\em iSRS} procedure to extract flares from simulated data (the same procedure used to extract
flares from real data).\\
Regarding case \emph{a}, we performed 100 simulations. We fitted the simulated waiting times distribution
with both an uniform, both with a multi-loghat distribution.
In fact the multi-loghat distribution approaches an uniform distribution for burst rates much larger than the flaring rate within a burst,
and we can successfully fit the uniformly distributed samples using a multi log-hat fitting function. 
Due to the telescope exposure variation with time to each source, the final sample of recognized flares could mimic a multi loghat distribution of waiting times.
We obtained for case \emph{a} (uniformly distributed flares) that the absolute variation of the Cash estimator from uniform to multi-loghat
fitting function is $\lvert\Delta C\rvert\ < 3$.
We already found for real data that the variation of the Cash estimator from uniform to multi-loghat fitting distribution is
$\Delta C\ = -105.8$. Thence the hypothesis that (for real data) uniformly distributed (in times) flares could mimics
a multi-loghat like distribution of waiting times (due to exposure variation with time) is rejected.
We found, instead, that fitting the multi-loghat function to the waiting time distributions obtained simulating case \emph{b} and \emph{c}, the fitted bursting rate and duration corresponded to the simulated bursting rate, and duration
of generated samples.
So, the fitting procedure and results obtained fitting the multi-loghat to the data is validated with simulations.\\
We note that for case \emph{c} the pile-up effect is reduced with respect to case \emph{b}:
For case \emph{c} flares within the same burst are highly correlated, because have all the same Lorentz and Doppler factor.\\
From  the detailed simulations we obtained that temporal resolving power reduces to $\frac{1}{2}$ at $\Delta_t$=  4d, 2d, 1.5d
respectively for case \emph{b},
case \emph{c}$_{cooling}$ (cooling time prevails), case \emph{c}$_{lct}$ (light crossing time prevails).\\
The detailed modelling cannot reproduce exactly the same number of detected flares, but we obtained that the revealed number of
simulated flares for each source was at most within a factor $\sim 2$ with respect to real data.
The comparison of the detailed simulation (case \emph{c}$_{lct}$) with the data is shown in Figure \ref{fig_data_vs_logstep_vs_unif}.
%
The comparison was performed excluding CTA 102 from the 100 MeV sample.
The waiting times obtained from the simulated time series reproduce the behaviour of data. Some discrepancy remains for short waiting times ($\Delta_t<3 d$):
In fact, with the detailed modelling, the pile-up effect shows up, suppressing the statistics for short waiting times.
Moreover, the  model does not try to reproduce the fast component acting on short timescales.\\  
To summarize, a set of superimposing bursts of activity is  able to reproduce the data. There is a suppression
of statistic for short waiting times.
Depending on the physical scenario, this suppression starts at $\Delta_t$ $\sim$1.5-4.
The observed waiting time distribution for all the sources (CTA 102 excluded) reported in fig \ref{fig_waiting_times} apparently does not show suppression.
We guess that the fast component compensates the pile-up suppression.\\
%
  We could expect that gamma-ray FSRQs selected in this paper should have a broad distribution of the jet axis with respect
 to the line of sight, affecting the observed peak flux, flare duration, and waiting times between flares.
 As a results the accumulated distribution of waiting times should be broad too, due to the different viewing angles among sources,
 possibly washing out the multi-loghat time distribution proposed to fit the results. 
 Instead, the flare luminosity dependency upon the doppler factor strongly mitigate the effect for flux limited samples:
 In fact
 flaring emitting regions, moving along a trajectory (that we identify with a flare direction)
 which is at large angle with respect to the line of sight, usually remain
 undetected in gamma-ray, because of the dependency of the flaring luminosity L$_{\gamma}$ upon the doppler factor
 (L$_{\gamma} \propto \delta^{4+2\alpha}$ for the external Compton, L$_{\gamma} \propto \delta^{3+\alpha}$ for the synchrotron emission,
 see , e.g., \citealt{dermer1997}). 
 This  suppression effect combines with the low probability to observe flares from electrons with bulk motion
 aligned with the line of sight.
As a result, viewing angles distribution for gamma-ray detected flares
 should be peaked
 at an intermediate value between 0 and the critical angle  $\Theta_{cr}=\frac{1}{\Gamma_{bulk}}$.
The distribution  of viewing angles of gamma-ray detected flares (obtained simulating the simplified scheme
{\emph c}$_{cooling}$ where the bulk direction of the flare emitting electrons coincides with the knot direction)
is reported in figure \ref{fig_knot_viewing_simulated} (top panel).
 This circumstance strongly reduces the broadening of the accumulated waiting time distribution.
 Suppression was investigated for flux-limited radio samples too, with jet viewing angle distribution peaking
 at about half of the critical angle  \citep[see, e.g., ][]{vermeulen1994,lister1997}. 
 Further suppression effect for gamma-ray detected blazars was also found in \citealp{savolainen2010}, and in
 \citealp{pushkarev2017}, see their Figure 15, reporting the distribution of
 jet viewing angles of the Mojave Radio sample for Fermi-LAT detected and for Fermi-LAT not detected AGNs.
 Our sample shows strong suppression at large viewing angles, because it is obtained starting from Fermi-LAT detected FSRQs
 for which we revealed flaring activity. Thence the accumulated distribution of waiting times between flares that we studied 
 here is obtained for sources with a narrow angle between the line of sight and the jet axis. 
 The simulated cases {\emph c}$_{lct}$ and {c}$_{cooling}$ show this suppression effect, and take into account for the
 non-monochromatic distribution of the bulk doppler factor of travelling knots \citep[see][]{lister2019}.\\  
 In the case of uniformly distributed flares within a burst of activity, we expect for each source a waiting time
 distribution $\rho(log(\Delta_t))e^{-\Delta_t}$ with a flat tail for short waiting times. As a result, the accumulated distribution
 (obtained accumulating waiting times for all the sources) 
 should still have a flat tail for short waiting times, irrespective of the differences of burst rate and duration among sources.
\\
\subsection{constraints on gamma-ray emission models}
\citet{jorstad2017}, \citet{casadio2019} argued that gamma-ray flares could be associated with superluminal moving features
crossing stationary features along the jet. If this is the case, the burst activity (that we revealed from waiting times between activity peaks)
could be associated with plasma streams relativistically moving along the jet, and crossing a steady feature along their path.
Assuming a stream with $\Gamma_{stream}$= 10, its typical length (in its reference system)
measured along the jet is $l'_{stream}=\tau_{burst}\beta_{stream}c\Gamma_{stream}\ \sim$2 pc.\\

Alternatively, we could assume that a travelling perturbation or a travelling knot is responsible for the bursting activity.
We could assume that in this case the direction of accelerated particles is highly correlated with the direction of the
travelling knot (corresponding to case \emph{c}).
We can correlate the duration of the observed burst with the travelled length $l_{knot}$ of the superluminal feature:
\begin{equation}
  \tau_{burst}\ =\frac{l_{knot}}{\beta_{knot}c}(1-\beta_{knot}cos\theta_{view})\ =\ \frac{l_{knot}}{\beta_{knot}c\Gamma_{knot}\delta_{knot}}
  \label{eq:tauburst}
\end{equation}
In this scenario the viewing angle of the jet axis varies from source to source, and the Lorentz factor of moving knots
is not monochromatic.
Adopting  the proposed case {\emph c}, the distribution of $\Gamma_{knot}\delta_{knot}$ for gamma-ray
detected flares can be evaluated with simulations. It is reported in figure \ref{fig_knot_viewing_simulated} (bottom panel)
assuming the mean Lorentz factor for the moving feature is $\Gamma_{knot}$=10.
We obtained a distribution of $log(\delta_{knot}\Gamma_{knot})$ with mean value  $\sim2.26$ and 
root mean square broadening $\sim27\%$; thence $l_{knot}$ is 30-50 pc. 
 We underline that, on account of Equation \ref{eq:tauburst} and of the spread of $log(\delta_{knot}\Gamma_{knot})$,
the viewing angle variation from source to source,
and the distribution of $\Gamma_{bulk}$  could be, at least partially,
responsible for the spread  of $\tau_{burst}$ reported in Table \ref{tab:fit}
(parameter $\sigma_{log(\tau)}$ of the multi-loghat distribution).\\
We can compare the achieved results with the VLBA observations reported in \citealt{jorstad2017}.
In their Table 7, they report the fading time $\tau_{43\ GHz}$ for moving features observed at 43 GHz.
\begin{figure*}
\includegraphics[width=8 cm]{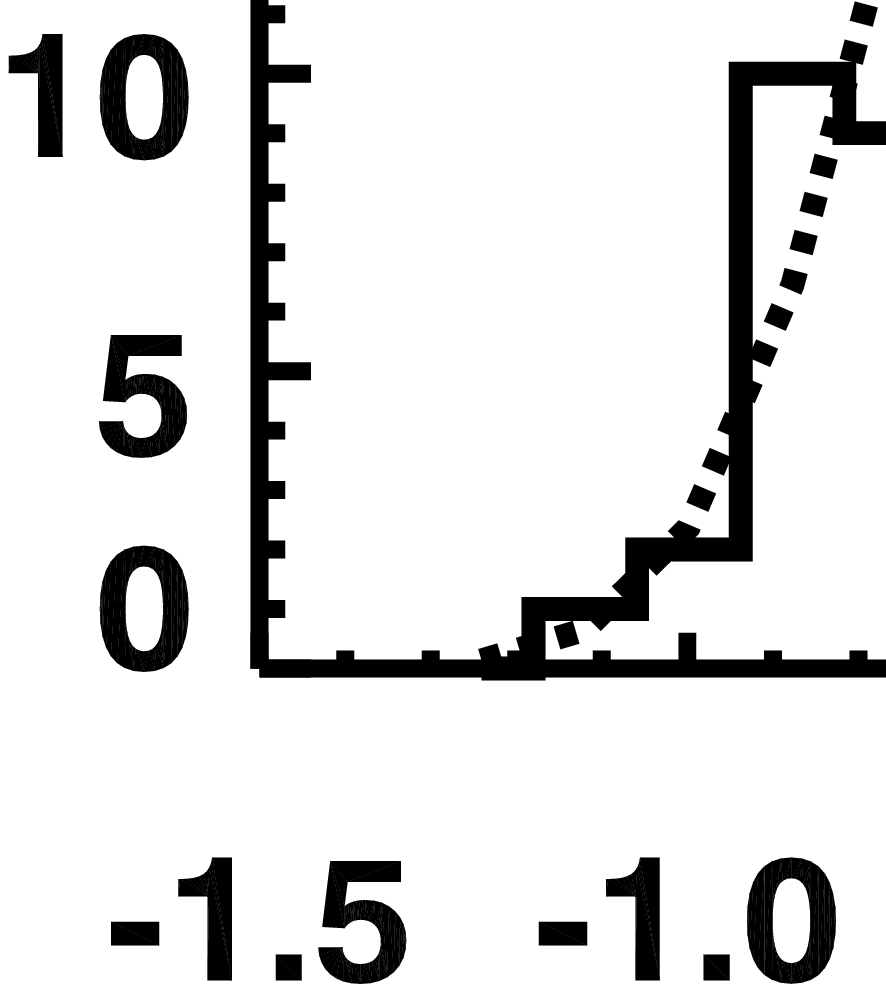} 
\includegraphics[width=8 cm]{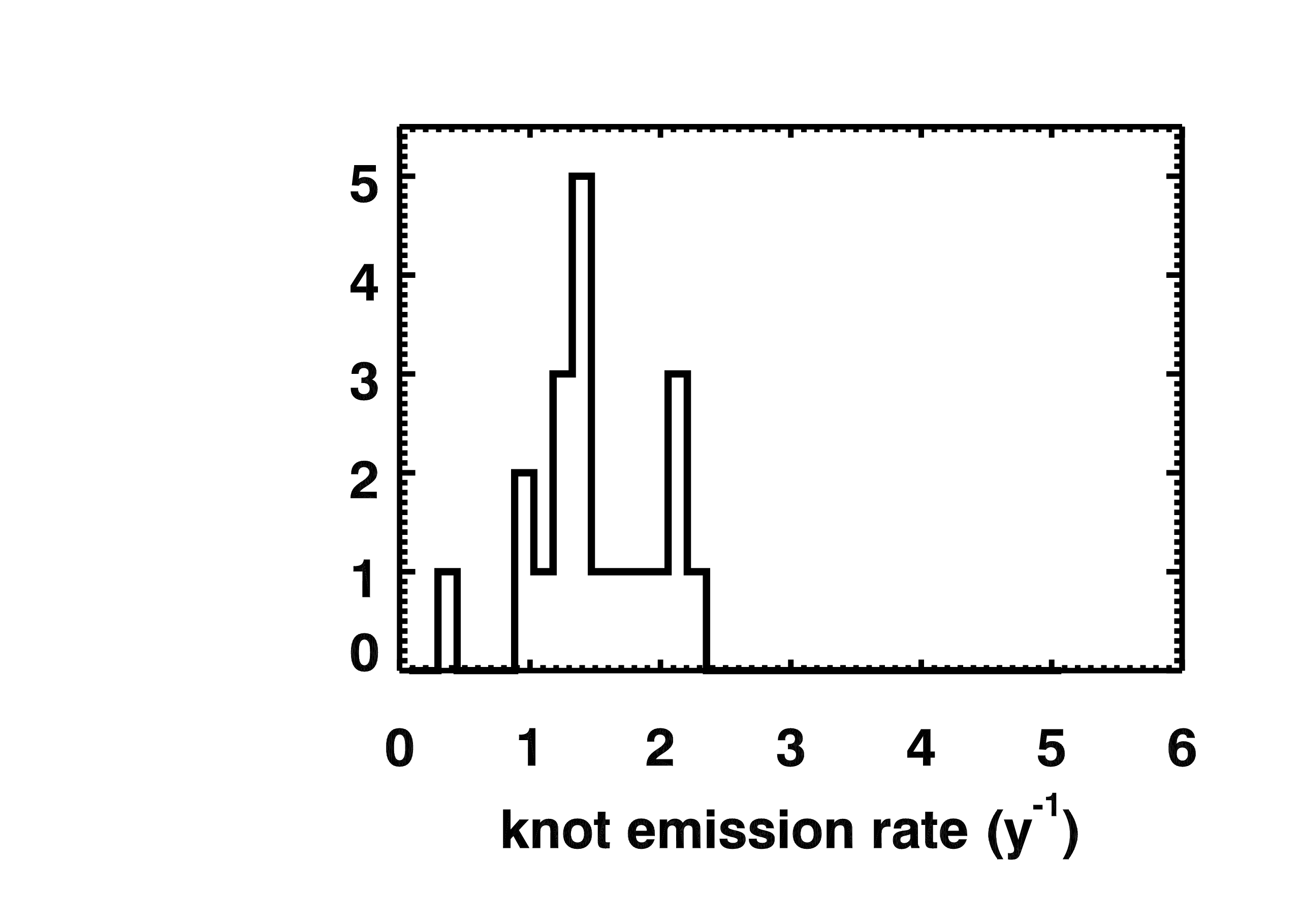} \\
\caption{Left panel: distribution of fading time of superluminal features observed at 43 GHz \citep[reported in Table 4 of][]{jorstad2017} (log scale).
Right panel: distribution of resolved superluminal features ejection rate at 43 GHz \citep[from Table 5 of][]{jorstad2017}. }
\label{fig_knot_43ghz}
\end{figure*}
We report in Figure \ref{fig_knot_43ghz} the $\tau_{43\ GHz}$ distribution and emission rate
for the resolved superluminal features observed at 43 GHz
\citep[see Figures 4 in ][]{jorstad2017}.
The mean value for $log(\tau_{43\ GHz} /y)$ of the fitting logarithmic distribution is $-0.50\pm0.02$,
and the standard deviation is 0.22$\pm0.02$ respectively.
It is tempting to associate the distribution of fading time of moving features, with the distribution
of the duration of bursts of flares found in this paper and reported in Table \ref{tab:fit}.
The comparison is reported in Figure \ref{fig_ulen_vs_rate}:
Our results agree with  the knot emission rate and with the fading time (within a factor two) of knots
observed at 43 GHz \citep{jorstad2017}.\\
As the knot travels away from the SMBH,
we expect a significant reduction of peak flux of the emitted flares. In fact
both the external magnetic field $B$, and the energy density of the external radiation field $u_{ext}$
decrease along the travelled path ($B\propto\frac{1}{d}$, and $u_{ext}\propto\frac{1}{d^2}$, with $d$
the distance of the emitting zone from the SMBH). Thence the peak flux of flares generated along the travelled path
follows the decreasing trend with $d$ in optical
(synchrotron emission) and  in gamma-ray (assuming EC emission).
We note indeed that the magnetic field reduction with the distance from the SMBH (and electron cooling)
should affect also
the fading time evaluated from radio luminosity of superluminal
features \citep{jorstad2017}. 
\citealt{homan2015} showed that radio features accelerate for deprojected distances up to $\sim$100 pc from the central SMBH.  
If flares are emitted while knots still experience acceleration, the enhancement of $\Gamma_{knot}$ could partially
compensate the reduction of energy density of the external photon field.\\

Finally, in magnetic reconnection scenario \citep[see, e.g.,][]{giannios2013}, reconnection events can be triggered by
magnetic instabilities such as kink instabilities \citep{begelman1998,spruit2001}, or
magnetic field inversions at the base of the jet \citep{giannios2019}.
We speculate that bursting activity we reveal could be generated by intermittent magnetic instabilities or by
intermittent flux of plasma:
instabilities phases of the magnetic structure of the jet, or the plasma injection to the reconnection sites should have a typical timescale of $\sim$0.6y.\\
\citet{sironi2016} obtained that plasmoids can be produced continuously (plasmoids chain) if fresh plasma is added to the system.
Due to the limited flare resolving power of the FERMI-LAT, we are rarely able to resolve plasmoid emission from the envelope \citep[see, e.g.,][]{christie2019}. 
Thence, the activity peaks we observe in gamma-ray correspond to the envelope emission from reconnection events.
If this is the case, reconnection  events
must be uniformly produced in time during the development of magnetic instabilities.
Moreover, we could associate the fast component to the rare cases for which we are able to resolve plasmoids within the envelope (when the jet axis and
reconnection layer are almost aligned to the line of sight).
In this physical scenario, the orientation of the reconnection layer differs for each reconnection event
(such as in simulated case \emph{b}), but plasmoids from the same reconnection event emit along the same direction. \\
\subsection{The Fast component and the gamma-ray data of CTA 102}
We observed a few tens of short waiting times (for waiting times $<$ 1d),
mainly detected for CTA 102. The typical timescale of this component is $\sim$1 hour.
The detailed simulations, performed taking into account for the FERMI-LAT exposure to the source,
and realistic duration of flaring activity show that 
temporal pile-up of close-by flares acts for waiting times $<$ 1.5-4d.
We did not try to study the distribution of the fast component in more details with the simple
fitting provided here. In fact pile-up plays a major role. Moreover
the exposure to sources cannot be considered continuous for time periods below several hours.
A study could be performed applying detailed modelling for particle acceleration and gamma-ray emission
to the simulating chain.\\
A fast, white component, with similar timescale was found in optical for two BL Lac objects \citep{raiteri2021a,raiteri2021b}.\\
Interestingly, the fast component shows up in CTA 102 gamma-ray data for the time interval for which the
superluminal component K1 
crosses the C1 stationary feature of the jet, and for this event,
the scenario of acceleration of turbulent cells from a recollimation shock was
proposed \citep{casadio2019}.\\
\citet{larionov2017} interpreted the CTA 102 period of activity differently:
They argued that a superluminal knot changing its own orientation and moving along the line of
sight from the end of 2016 to the beginning of 2017 (during the period of conspicuous detection of the fast component)
can produce the observed flux and polarization trend in optical and radio. We
note that the change in orientation could explain the quadratic
correlation observed in optical and gamma-ray \citep{larionov2017}: In fact,
it corresponds to a change in Doppler factor and then to a change in gamma-ray
luminosity $L_{gamma-ray}\propto\delta^{4+2\alpha}$ (with $\alpha$ the gamma-ray spectral
index), and to a change in optical luminosity $L_{optical}\propto\delta^{3+\alpha}$, 
according to \citet{dermer1997} for EC and synchrotron emission respectively.
The weak polarization observed on 23-Dec-2016
reported in \citet{casadio2019} for the K1 superluminal feature, witnesses for an aligned
direction of the K1 component toward the observer.\\
The change in orientation scenario causes less than a factor 2 reduction in
waiting times of gamma-ray, but it is responsible for a large increase of flux, and thence
the activity peaks can be temporally resolved with the FERMI-LAT
(we already discussed the net effect in the results for CTA 102). Thence,
at least for a short period of time, we were able to resolve in gamma-ray the
fast component flaring activity, due to turbulence, or to the plasmoid chain in reconnection scenario.
\\

\section{Summary}
In this paper we found that gamma-ray activity peaks of FSRQs are produced in bursts. The average
duration of a single burst is $\sim$0.6y, and the average
burst rate is $\sim$1.3y$^{-1}$. These are averaged values obtained from the whole analysed sample.
The temporal distribution of activity peaks within each burst can be approximated with an uniform distribution.\\
Our overall distribution of waiting times between activity peaks shows
a statistically relevant fast component (with $\Delta_t<$1d), that can be roughly modelled with an uniform distribution
with a typical waiting time of $\sim$1h. CTA 102 shows the large
majority of the fast population.
Once CTA 102 is removed from the whole sample, the cumulative waiting time distribution of all the  remaining sources
does not show clear evidence of the fast population.
Indeed, several cases of short waiting times, that can be ascribed to the fast component,
can be found in the waiting time distribution of single sources. We evaluated the pile-up effect on the waiting time sample,
and found that it must be relevant for waiting times shorter 1.5-4d, depending on the modelling.
We guess that the fast component and the pile-up compensate each other in the cumulative distribution of waiting times
of the remaining sources.  Thence the fast component should act for waiting times shorter than 1.5-4d.\\
It's worth mentioning that \citealp{raiteri2021a,raiteri2021b} found a statistically relevant white noise emerging for frequencies above $\sim0.9h^{-1}$ and $\sim3.6h^{-1}$
in the power spectral density analysis (PSD) of optical data of the BL Lac objects S4 0954+65 and S5 0716+714
respectively (observed with
TESS, Transiting Exoplanet survey Satellite, \citealt{ricker2015}, and WEBT\footnote{http://www.oato.inaf.it/blazars/webt/},
Whole Earth Blazar Telescope organization, \citealt{villata2002}).\\

Our finding can be applied to existing theoretical/phenomenological models: 
\citep{narayan2012,marscher2013,marscher2014} proposed that gamma-ray flares originate from the passage of turbulent plasma
inside a recollimation shock. In such a scenario, the burst of activity that we found could be associated with
the length of a superluminal plasma stream. We estimated a length of $\sim$2 pc in the stream reference frame
(assuming a Lorentz factor of the stream $\Gamma_{stream}=10$, and a characteristic recollimation shock length much shorter than
the stream length in the host galaxy frame).\\
Instead, if flares are generated along the path of a superluminal moving perturbation (or knot),  the average path length should
be 30-60 pc (assuming that the Lorentz factor of the superluminal feature is $\Gamma_{knot}$=10). This scenario
is not compatible with the $1/d^2$ scaling of the energy density of the magnetic and external photon field in leptonic models
(with $d$ the distance of the superluminal feature from the SMBH). The effect on emission of the $1/d^2$ scaling of
external fields could be partially mitigated if the superluminal feature accelerates along its path.\\ 
Finally the reconnection particle accelerating scenario \citep{giannios2013} divides the acceleration mechanism
in the generation of reconnection events,
and in the plasmoid-chain grown within each reconnection event \citep{sironi2016}.
In this scenario, we should associate the bursting activity
to the jet magnetic field instabilities (that generate reconnection events) or to the plasma injection to the reconnection sites,
and the fast component to the gamma-ray emission from the plasmoid chain. Thence the typical timescale of persistence
of a single instability (or of duration of the plasma injection) 
should be $\sim$0.6y, and the rate of generation of magnetic instabilities (or the rate of plasma injection) should be $\sim$1.3$y^{-1}$. Moreover the typical waiting times between gamma-ray detected
plasmoids should be $\sim$1h.\\

Several authors \citep[see, e.g., ][]{kelly2009,ivezic2013} found that Quasars optical variability can be described with
a damped random walk, with damping timescale $\tau_{damping}$ of the order of several hundreds of days. \citealp{kelly2009} also
observed that Radio Loud Quasars show an excess optical variability for timescales below 1 d, with a white noise PSD.
Recently \citealp{burke2021} found that the damping timescale found in optical correlates with SMBH mass in AGNs:
$\tau_{damping}\sim$ 110-260d for  masses in the range of 10$^8$-10$^9M_{\odot}$.\\
In our analysis we found similar timescales. This similarity suggests that physical mechanism responsible for
optical variability on long timescales
could be correlated with the gamma-ray emission process responsible for burst activity of FSRQs.\\

\section*{Appendix A: temporal distributions for flaring times adopted for fittings:}
\begin{figure}
\includegraphics[width=7 cm]{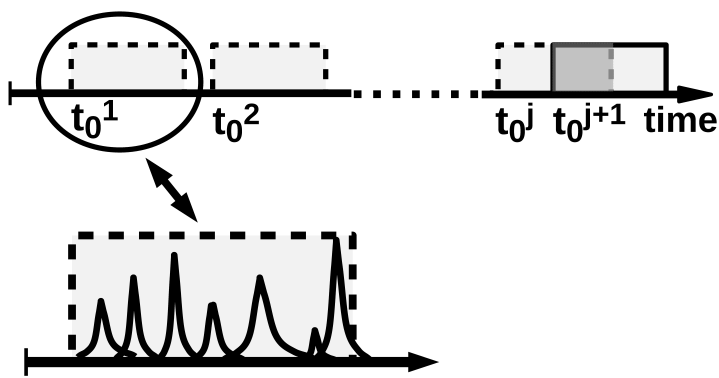} \\
\caption{example of burst activity (multi-hat distribution): each gray rectangle represents a burst of activity. Flares are produced with an uniform distribution inside the bursts of activity. The burst starting time $t_0^j$ is uniformly
distributed along the observing period, the duration of all the bursts is the same. Bursts can overlap.}
\label{fig_burst}
\end{figure}
In this study we make use of several distributions of flaring times:\\
\begin{itemize}
\item
An uniform distribution, characterized by the total number of flares $N_{flares}$.\\
\item
an overlapping set of uniformly distributed bursts of activity 
with all bursts of the same duration $\tau_{burst}$ (called multi-hat, see fig. \ref{fig_burst}).
Flaring probability is uniform within the burst.\\
This distribution is characterized by the burst rate parameter $r_{burst}$, by $\tau_{burst}$, and $N_{flares}$;\\
\item
an overlapping set of uniformly distributed bursts of activity 
with $\tau_{burst}$  distributed with a log normal distribution (called multi-loghat).
Flaring probability is uniform within the burst.\\
This distribution is characterized by the burst rate parameter $r_{burst}$, by $N_{flares}$ and the parameters of the log-normal
distribution ($m_{log(\tau)}$ and $\sigma_{log(\tau)}$).\\
\item
an overlapping set of bursts of activity 
with all bursts of the same duration $\tau_{burst}$. The distribution of flares within each
burst follows a power-law distribution $(\rho_{pow}$, reported in Equation \ref{eq:tpow}),
\begin{equation}
  \rho_{pow}\ \propto\ \frac{1}{|t-t_0|^{1-\frac{1}{\eta}}}\ \ \ \ \ with\ \ \  \eta\ >\ 0
  \label{eq:tpow}
\end{equation}
around a pole $t_0$ conventionally placed at the end of the bursting period.\\
This distribution is characterized by the burst rate parameter $r_{burst}$, by $N_{flares}$, $\eta$ coefficient, and $\tau_{burst}$.\\
This distribution is called multi-pow.\\ 
\item
an overlapping set of bursts of activity, 
with  bursts of duration $\tau_{burst}$ distributed with a log normal distribution. The distribution of flares within each
burst follows a power-law distribution $(\rho_{pow}$, reported in \ref{eq:tpow}),
around a pole $t_0$ conventionally placed at the end of the bursting period.\\
This distribution is characterized by the burst rate parameter $r_{burst}$, by $N_{flares}$, $\eta$ coefficient, and the parameters of the log-normal
distribution ($m_{log(\tau)}$ and $\sigma_{log(\tau)}$).\\
This distribution is called multi-pow-smooth.\\ 
\item
A composite model obtained adding a poissonian process to the multi-loghat fitting function:
The event time of a fraction R$_{fast\ observed}$ of the events,  is extracted at a time distance $\Delta_t$ from an event of the multi-loghat distribution.
$\Delta_t$ is extracted with an exponential distribution $\rho(\Delta_t)=\frac{e^\frac{-\Delta_t}{\tau_{fast}}}{\tau_{fast}}$.\\

This distribution is characterized by the burst rate parameter $r_{burst}$, by $N_{flares}$, by the parameters of the log-normal
distribution ($m_{log(\tau)}$ and $\sigma_{log(\tau)}$), and by $R_{fast\ observed}$, $\tau_{fast}$.\\
This distribution is called multi-loghat + poissonian.\\
\item
A composite model obtained adding a poissonian process to the multi-pow fitting function:
The event time of a fraction R$_{fast\ observed}$ of the events,  is extracted at a time distance $\Delta_t$ from an event of the multi-pow distribution.
$\Delta_t$ is extracted with an exponential distribution $\rho(\Delta_t)=\frac{e^\frac{-\Delta_t}{\tau_{fast}}}{\tau_{fast}}$.\\

This distribution is characterized by the burst rate parameter $r_{burst}$, by $N_{flares}$, $\eta$ coefficient, and $\tau_{burst}$ of the multi-pow
distribution, and by $R_{fast\ observed}$, $\tau_{fast}$.\\
This distribution is called multi-pow + poissonian.\\
\end{itemize}
Fitting method is explained in appendix B.\\
\section*{Appendix B: fitting method}
%
%
The realization of the fitting distribution is obtained  with a montecarlo method: it is the simulated distribution of waiting times.\\
It is assumed that waiting times distribution is the same for all the sources,
and that only the observed number of flares can vary from source to source. \\
For each source, we extract the number of emitted bursts within the observing period, their starting time, and length.
Finally, starting from the observed number of flares for the chosen source, for each flare we extract the burst to which the flare belongs,
and the flare starting time within the burst.
This last choice corresponds to the assumption that peak flux and duration of each flare does not depend on emission time within a burst.
For each source we perform one thousand of simulations.\\
We adopted the binned Cash statistic \citep{cash1979} to fit model to the data.
There is a chance that a null number of bursts is extracted for a given source. In this case, the simulation cannot reproduce the
observed number of flares. So, for each source, the probability to observe a given configuration is obtained by the product of:
the probability that the extracted number of burst is not null ($1-P_{null}^{src}$)  and the usual term discussed in \citet{cash1979}.
Thence we have to add an element for each source to the Cash estimator; this element is $-2ln(1-P_{null}^{src})$.\\
\section*{Appendix C: temporal pile-up effect on the waiting times sample}
\begin{figure*}
\includegraphics[width=8 cm]{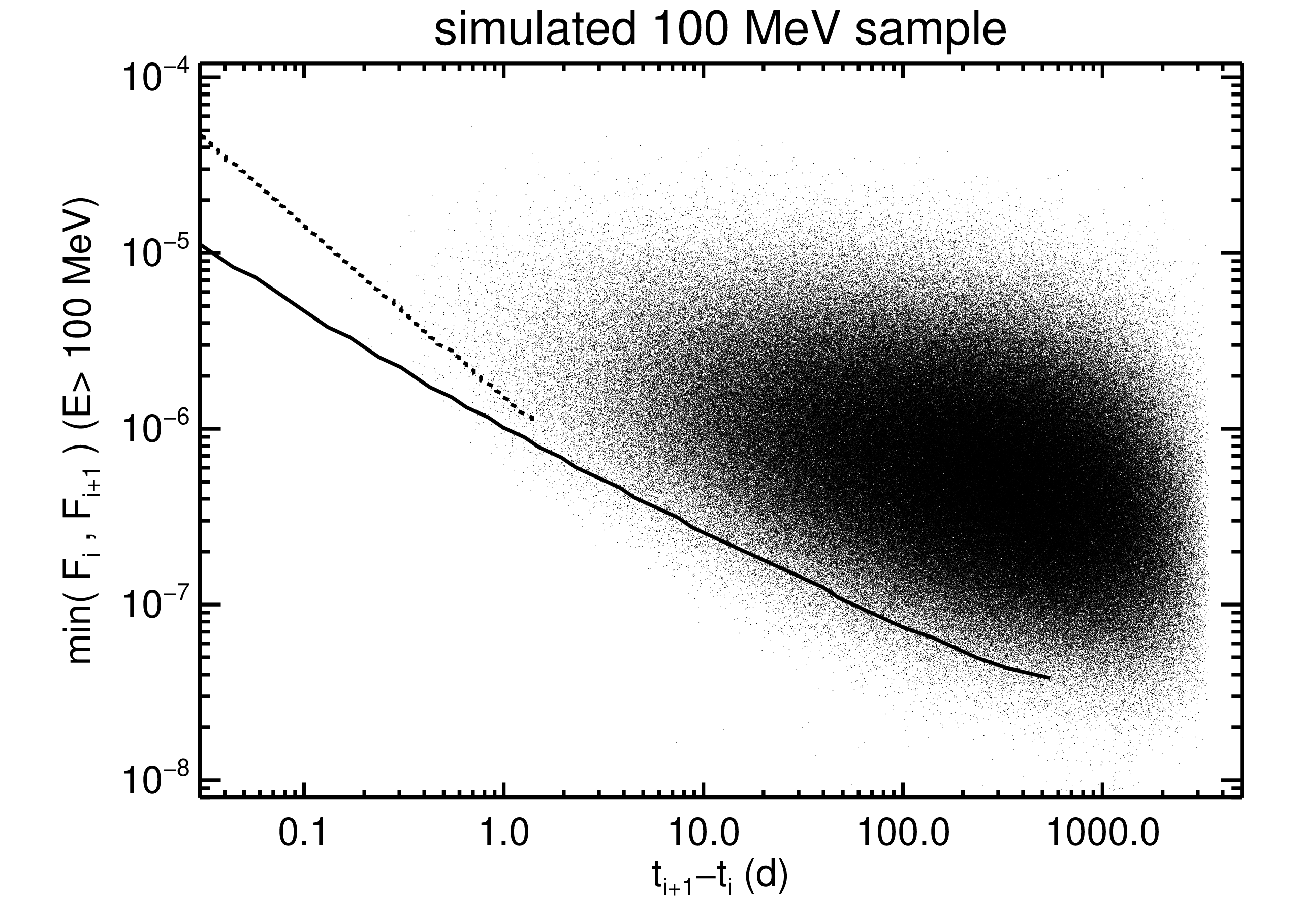} \includegraphics[width=8 cm]{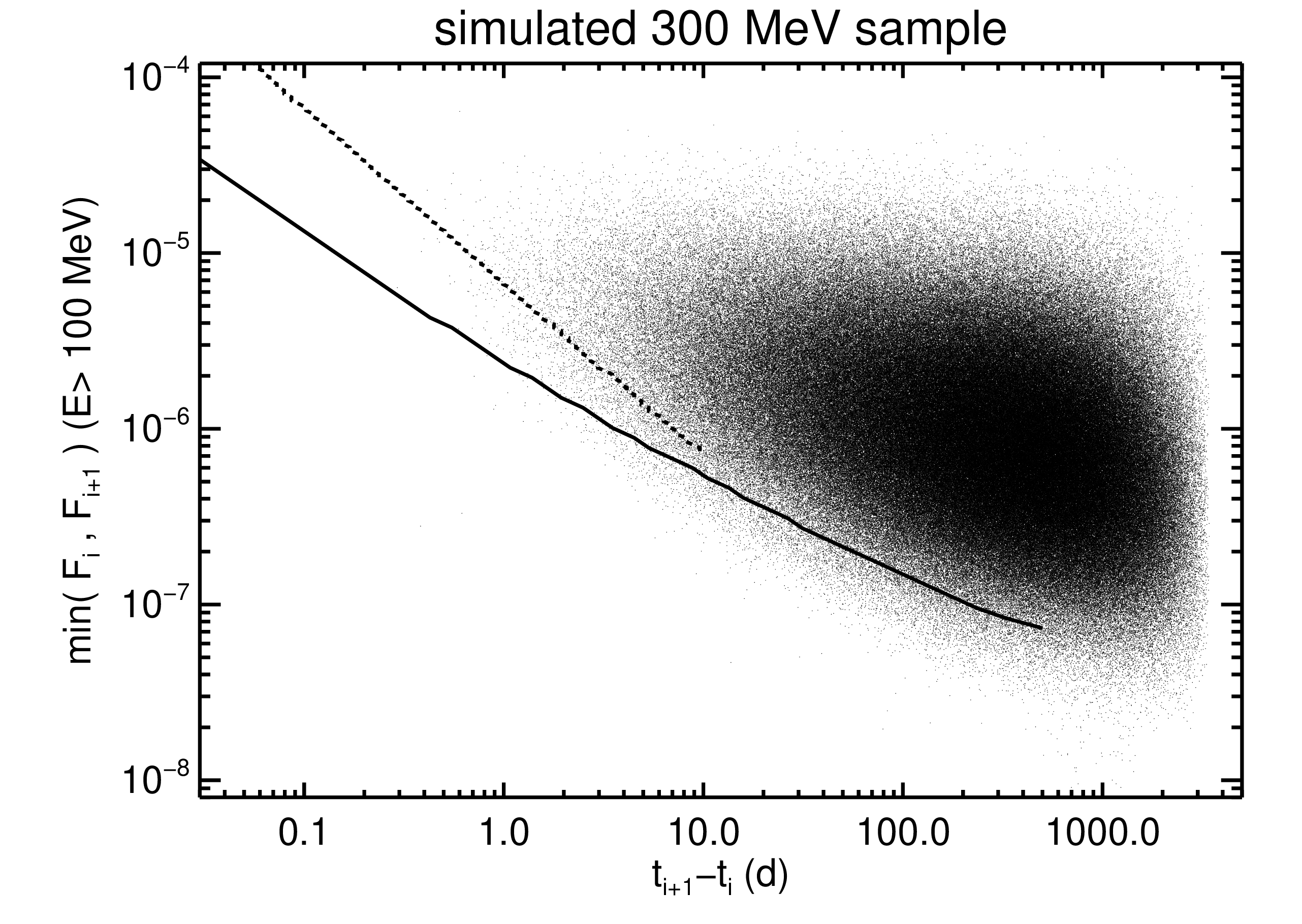} \\
\caption{distribution of $min(F_i,F_{i+1})$ vs $t_{i+1}\ -\ t_i$ for
  the simulated 100 and 300 MeV samples, left and right plot respectively.
  For the 300 MeV sample, the flux was scaled to match the flux for gamma-ray above 100 MeV
  (a power-law flux with photon index of 2.23 is assumed).
  solid line is the sensitivity limit for a flare from the bright source 3C 454.3 with FWHM twice
  the waiting times $t_{i+1}\ -\ t_i$.  Dashed line is the temporal resolving power.
}
\label{fig_pileup}
\end{figure*}

Often authors study crowded periods of activity of blazars
trying to resolve single flares with a
fitting strategy \citep[see, e.g.,][]{abdo2011,meyer2019}; a predefined flare temporal shape for flares is adopted.
Often the shape is parameterized as in equation 7 in \citet{abdo2010}.\\
In this paper we do not make use
of this method. Only statistically detected peaks with the \emph{iSRS} method
belongs to the studied sample.
During crowded periods of activity, temporal pile-up of flares
results in a reduction of the number of detected activity peaks.\\
Pile-up effect was already addressed in appendix B of \citet{pacciani2018}.
Here we compare that findings with simulated samples.\\
We simulated flares with a multi-loghat temporal distribution of flaring times.
Once the simulated temporal series were obtained for each source, we performed the
\emph{iSRS} procedure to obtain the flaring peak time, luminosity and duration.\\
The results for the simulated 100 and 300 MeV samples are summarized in figure \ref{fig_pileup}.
We plotted the waiting times versus the minimum among peak fluxes of  two consecutive resolved flares
($min(F_i,F_{i+1})$, where $F_i$ and $F_{i+i}$ are the measured peak fluxes of consecutive flares).
to facilitate the comparison, in both the plots the reported flux is evaluated for E>100 MeV (a power-law index
is assumed with index 2.23). The comparison of the two plot reveals that the resolving power is larger for the 100 MeV
sample: roughly speaking, the larger statistics allows for an easier separation of close-by flares.\\
In the same figure, the sensitivity limit is reported for the detection of a flare with temporal
FWHM which is the double of the waiting time, and flux corresponding to $min(F_i,F_{i+1})$.
A curve corresponding to the resolving power of consecutive flares is reported
too \citep[see][for details]{pacciani2018}.\\
\begin{acknowledgements}
LP thanks IAPS-INAF for support on funds Ricerca di Base
F.O. 1.05.01.01, and contribution from the grant INAF Main Stream project
High-energy extragalactic astrophysics: toward the Cherenkov Telescope Array,
F.O. 1.05.01.86.26.
We acknowledge all Agencies and Institutes supporting the Fermi-LAT operations
and the Scientific Analysis Tools. L.P. is grateful to Fabrizio Tavecchio for
useful suggestions.
\end{acknowledgements}

\end{document}